\documentclass[twocolumn]{aastex62}
\usepackage[utf8]{inputenc}

\usepackage{amsmath}
\usepackage{comment}

\newcommand{\mmin}{m_\mathrm{min}}
\newcommand{\mmax}{m_\mathrm{max}}

\graphicspath{{./}{figures/}}

\newcommand{\pp}[1]{\textcolor{black}{#1}}
\newcommand{\daniel}[1]{}

\newcommand{\julymassppop}{38.9^{+7.3}_{-4.5} \ M_\odot}
\newcommand{\julymasspLI}{51.2^{+16.2}_{-11.0}
 \ M_\odot}
\newcommand{\julyqpop}{0.95^{+0.05}_{-0.18}}
\newcommand{\julyqLI}{0.63^{+0.32}_{-0.26}}

\newcommand{\julymppdpA}{69}
\newcommand{\julymppdpAexc}{86}
\newcommand{\julymppdpB}{59}

\newcommand{\KSpB}{84\%}
\newcommand{\ADpB}{79\%}
\newcommand{\KSpAexc}{69\%}
\newcommand{\ADpAexc}{71\%}
\newcommand{\KSpA}{87\%}
\newcommand{\ADpA}{82\%}

\begin{document}

\title{The Most Massive Binary Black Hole Detections and the Identification of Population Outliers}

\author{Maya Fishbach}
\affiliation{{Department of Astronomy and Astrophysics, University of Chicago, Chicago, IL 60637, USA}}
\email{mfishbach@uchicago.edu}

\author{Will M. Farr}
\affiliation{Center for Computational Astrophysics, Flatiron Institute, 162 Fifth Avenue, New York, NY 10010, USA}
\affiliation{Department of Physics and Astronomy, Stony Brook University, Stony Brook, NY 11794-3800, USA}
\email{will.farr@stonybrook.edu}

\author{Daniel E. Holz}
\affiliation{Enrico Fermi Institute, Department of Physics, Department of Astronomy and Astrophysics,\\and Kavli Institute for Cosmological Physics, University of Chicago, Chicago, IL 60637, USA}
\email{holz@uchicago.edu}

\begin{abstract}
Advanced LIGO and Virgo detected ten binary black holes (BBHs) in their first two observing runs (O1 and O2). Analysis of these events found strong evidence for a dearth of BBHs with component masses greater than $\sim45 \ M_\odot$, as would be expected from a pair-instability mass gap. Meanwhile, a standalone analysis of the merger GW170729 found its primary mass $m_1 = {51.2^{+16.2}_{-11.0} \ M_\odot}$, with the majority of its posterior support at $m_1 > 45 \ M_\odot$. Although this appears to be in contradiction with the existence of a limit at $\sim45\ M_\odot$, we argue that individual events cannot be evaluated without reference to the entire population. When GW170729 is analyzed jointly with the rest of the detections, as part of a full hierarchical population analysis, its inferred primary mass tightens considerably, to $m_1 = {38.9^{+7.3}_{-4.5} \ M_\odot}$. For a large sample of events in the presence of noise, apparent outliers in the detected distribution are inevitable, even if the underlying population forbids outliers. We discuss methods of distinguishing between statistical fluctuations and population outliers using posterior predictive tests. Applying these tests to the primary mass distribution in O1 and O2, we find that the ten detections are consistent with even the simplest power-law plus maximum-mass model considered by the LVC. This supports the claim that GW170729 is not a population outlier. We also provide non-parametric constraints on the rate of high-mass mergers and conservatively bound the rate of mergers with $m_1 > 45 \ M_\odot$ at $2.8^{+5.4}_{-2.0}\%$ of the total merger rate.  After 100 detections like those of O1 and O2 from a population with a maximum primary mass of $45 \, M_\odot$, it would be common for the most massive system to have an observed maximum-likelihood mass $m_1 \gtrsim 70 \, M_\odot$.
\end{abstract}

\section{Introduction}
A major goal of gravitational-wave (GW) astronomy is to learn about the formation and evolutionary mechanisms of binary black hole (BBH) mergers, such as those detected by Advanced LIGO~\citep{TheLIGOScientific:2014jea} and Virgo~\citep{TheVirgo:2014hva}. There are many proposed formation channels for BBHs, including isolated evolution~\citep{2015ApJ...806..263D, 2016Natur.534..512B,  2016ApJ...824L..10W, 2016MNRAS.462.3302E, 2016ApJ...819..108B,2017NatCo...814906S,2018MNRAS.481.1908K, 2019MNRAS.485..889S}, dynamical formation~\citep{2016MNRAS.459.3432M, 2016PASA...33...36H, 2016ApJ...824L...8R, 2017MNRAS.464L..36A, 2017ApJ...836L..26C, 2018PhRvL.120o1101R, 2018PhRvD..97j3014S, 2019ApJ...871...91Z, 2019arXiv190100863D}, and primordial origin~\citep{2016PhRvL.116t1301B,2017JPhCS.840a2032G}, and if several formation channels are active at once, the population of merging BBHs may consist of distinct sub-populations. These sub-populations may differ in their shape of the mass distribution and spin distribution, as well as the merger rate (and its evolution with redshift).

Previous studies have explored methods of distinguishing between different formation channels using GW observations of BBHs, including fitting for the mixture fraction (or branching ratio) between various sub-populations~\citep{2015ApJ...810...58S,2016ApJ...832L...2R,2017ApJ...846...82Z,2017CQGra..34cLT01V,2017MNRAS.471.2801S,2019arXiv190511054B}. One proposed sub-population includes second-generation mergers, which occur when at least one of the component BHs in a binary is itself the product of a previous merger. Second-generation BHs are expected to have a characteristic distribution of dimensionless spin magnitudes that peaks at $a \sim 0.7$ and a mass distribution that extends into the ``pair-instability" mass gap starting at $\sim40$--$50 \ M_\odot$~\citep{2017ApJ...840L..24F,2017PhRvD..95l4046G,2018PhRvL.120o1101R}. Second-generation mergers are possible (and generally expected to occur) exclusively in dense stellar environments such as globular clusters; therefore, the existence of this population is an important discriminator between dynamical and isolated formation. An additional proposed sub-population consists of gravitationally-lensed GW signals, for which the lensing magnification causes a bias in the inferred luminosity distance and the unredshifted, source-frame masses if not properly accounted for~\citep{2014PhRvD..90f2003C}. Therefore, gravitationally-lensed events, even if they originate from the same formation channel as the unlensed events, would appear as a sub-population of erroneously high-mass, low-distance events\pp{~\citep{2017PhRvD..95d4011D,PhysRevD.97.023012,10.1093/mnras/sty2145,2018arXiv180205273B,2019ApJ...874L...2H}}.

The LVC detected ten BBHs in its first two observing runs~\citep{2018arXiv181112907T}. With these ten detections, \cite{2018arXiv181112940T} fit simple parameterized models to the mass, spin, and redshift distribution of the BBH population, assuming that all detections belong to the same population\footnote{We note that in addition to the LVC-published detections of~\cite{2018arXiv181112907T}, new BBH detections in the O1 and O2 data have been reported by \cite{IAS:O1}, \cite{IAS:O2} and \cite{Nitz:Catalog}}. The assumption of a single population was justified by a leave-one-out analysis, which shows explicitly that excluding GW170729, the most ``unusual" event (in terms of having the highest mass, spin, and distance), from the analysis does not significantly impact the inferred mass, spin, and redshift distributions at a level beyond the statistical uncertainties. \cite{2019arXiv190307813K} and~\cite{2019arXiv190306742C} also found that there is insufficient evidence to claim GW170729 as a population outlier by specifically comparing the hypothesis that it belongs to a population of second-generation, as opposed to first-generation, mergers based on the expected mass and spin distributions under the two scenarios.

In this paper we examine in more detail whether the assumed single-population mass distribution is a good fit to the data from the first ten detections, with a focus on the inferred ``maximum mass," or lower edge of the pair-instability mass gap. We argue that there are no convincing outliers among the O1 and O2 detections, implying that a single-component population model is sufficient to fit the data.
As the number of detections increases, we expect standard statistical fluctuations to produce individual events with significant posterior support inside the so-called upper mass gap; GW170729 may be an example of such a fluctuation. Additionally, we forecast the masses of future detections based on the population of BBHs from O1 and O2, and explore the masses that would be required to identify a BBH detection as a true population outlier. Such an outlier may belong to an alternate population consisting of, e.g., second-generation mergers, or otherwise indicate that the assumed population model provides an insufficient description of the data.
The methods discussed here can be extended to classify any outliers, including in the spin, mass-ratio, or redshift distribution.

The parametric models discussed above and used in~\citet{2018arXiv181112940T} are designed to incorporate a high-mass feature inspired by the predicted pair-instability mass-gap, whether it is a sharp cut-off to the power-law~\citep[Models A and B of][]{2018arXiv181112940T} or a Gaussian component (Model C). To constrain the rate of high-mass BBH mergers in a model-agnostic way, we apply the non-parametric method of~\citet{2017MNRAS.465.3254M} to the ten BBHs from O1 and O2. This method models the mass distribution as a binned histogram in the $m_1$--$m_2$ plane, with a smoothing prior on the bin heights. In contrast to the parametric models, the non-parametric fit a priori prefers smoothness over sharp features, providing a conservative upper limit on the rate of high-mass mergers.

The remainder of the paper is organized as follows. In Section~\ref{sec:popprior}, we review the hierarchical Bayesian framework of a population analysis and demonstrate how the population fit provides updated inference on the event-level parameters, yielding a tighter measurement on the masses of GW170729 in particular. In Section~\ref{sec:PPD}, we use posterior predictive distributions to evaluate the goodness-of-fit of a model to observations, focusing on high-mass outliers. In Section~\ref{sec:binned}, we apply the non-parametric histogram model of~\citet{2017MNRAS.465.3254M} to the ten LVC events and compare the inferred mass distribution to the parametric inference. We conclude in Section~\ref{sec:conclusion}, and additional analysis details are provided in the Appendix.

\section{Population Prior on Individual Events}
\label{sec:popprior}
A population analysis is concerned with fitting features that are common across the population members; in this case, BBH merger events. We assume that the event-level parameters $\{\theta_i\}$ (e.g. the component masses of the BBH event $i$) follow a probability distribution function $p_\mathrm{pop}\left(\theta_i \mid \Lambda \right)$, where $\Lambda$ are the population-level hyperparameters (e.g. the power-law slope of the primary mass distribution).
A hierarchical Bayesian analysis simultaneously fits for the event-level parameters $\theta_i$ and the population-level hyperparameters $\Lambda$ from the data~\citep{2010ApJ...725.2166H,2010PhRvD..81h4029M,2011ApJ...731..120M}.
In the presence of selection effects and measurement uncertainty, the joint posterior probability distribution of the event-level parameters $\{\theta_i\}$ and the population parameters $\Lambda$  given the data $\{d_i\}$ from $N$ independent events is given by~\citep{2004AIPC..735..195L,2019MNRAS.486.1086M}:
\begin{equation}
    \label{eq:jointlikelihood}
    p(\{\theta_i \}, \Lambda \mid \{ d_i \}) = \pi(\Lambda)\prod_{i=1}^{N} \frac{ \mathcal{L}(d_i \mid \theta_i)p_\mathrm{pop}(\theta_i \mid \Lambda)}{\int P_\mathrm{det}(\theta)p_\mathrm{pop}(\theta \mid \Lambda) \mathrm{d} \theta},
\end{equation}
where $\pi(\Lambda)$ is the prior on the population hyperparameters, $\mathcal{L}(d_i \mid \theta_i)$ is the event-level likelihood, and $P_\mathrm{det}(\theta)$ is the probability of detecting a piece of data from a merger with true parameters $\theta$, as discussed in detail below.

In a population analysis, we usually marginalize Eq.~\ref{eq:jointlikelihood} over the event-level parameters $\theta_i$
to recover the posterior of the population parameters $\Lambda$.
Meanwhile, when analyzing data from an individual event (``parameter estimation," or PE), the posterior on the event-level parameters,
\begin{equation}
p(\theta_i \mid d_i) \propto \mathcal{L}(d_i \mid \theta_i)\pi(\theta_i),
\end{equation}
is typically calculated using a default uninformative prior $\pi(\theta_i)$, rather than a population prior $p_\mathrm{pop}(\theta_i \mid \Lambda)$. The mass estimates in~\cite{2018arXiv181112907T}, for instance, are obtained using priors that are flat in detector-frame component masses.
Alternatively, one can marginalize Eq.~\ref{eq:jointlikelihood} over the population-level parameters $\Lambda$, and get a new posterior on the event-level parameters for each detection. Therefore, by calculating a joint posterior on the population-level and event-level parameters simultaneously, a hierarchical Bayesian analysis in effect replaces the default uninformative priors from PE with a population prior, yielding an informed posterior on the event-level parameters.

Since the first BBH detections, a variety of hierarchical analyses have been carried out to fit the BBH mass, spin, and redshift distributions to phenomenological parametric models~\citep{2016PhRvX...6d1015A,2017PhRvD..95j3010K,2017ApJ...851L..25F,2018ApJ...856..173T,2019MNRAS.484.4216R,2018ApJ...863L..41F,2019PhRvD.100d3012W}. In this work, we focus on the population analysis of~\citet{2018arXiv181112940T}; specifically the mass distribution, which~\citet{2018arXiv181112940T} fit according to three power-law based models: Models A, B, and C. Model A has two free parameters: the power-law slope for the primary mass distribution and the maximum mass, and fixes the minimum mass to 5 $M_\odot$ and the conditional distribution for the secondary mass to be flat between the minimum mass and $m_1$. Model B is a generalization of Model A with two additional parameters: the minimum mass and the power-law slope of the secondary mass (or equivalently, mass ratio) distribution as conditioned on the primary mass. Model C generalizes Model B by tapering the low- and high-mass ends of the mass function (as opposed to the sharp cutoffs of Models A and B) and allowing a high-mass Gaussian component on top of the power-law in primary mass. For each model, one can use the posterior samples for the population parameters $\Lambda$ to recover the population-informed posteriors for individual event parameters $\theta_i$ according to Eq.~\ref{eq:jointlikelihood}.\footnote{The posterior samples on the population hyper-parameters from~\citet{2018arXiv181112940T} are publicly available at~\url{https://dcc.ligo.org/LIGO-P1800324/public} and the PE samples for individual events from~\cite{2018arXiv181112907T} are available at \url{https://dcc.ligo.org/LIGO-P1800370/public}.}

\begin{figure*}
    \centering
    \includegraphics[width=\textwidth]{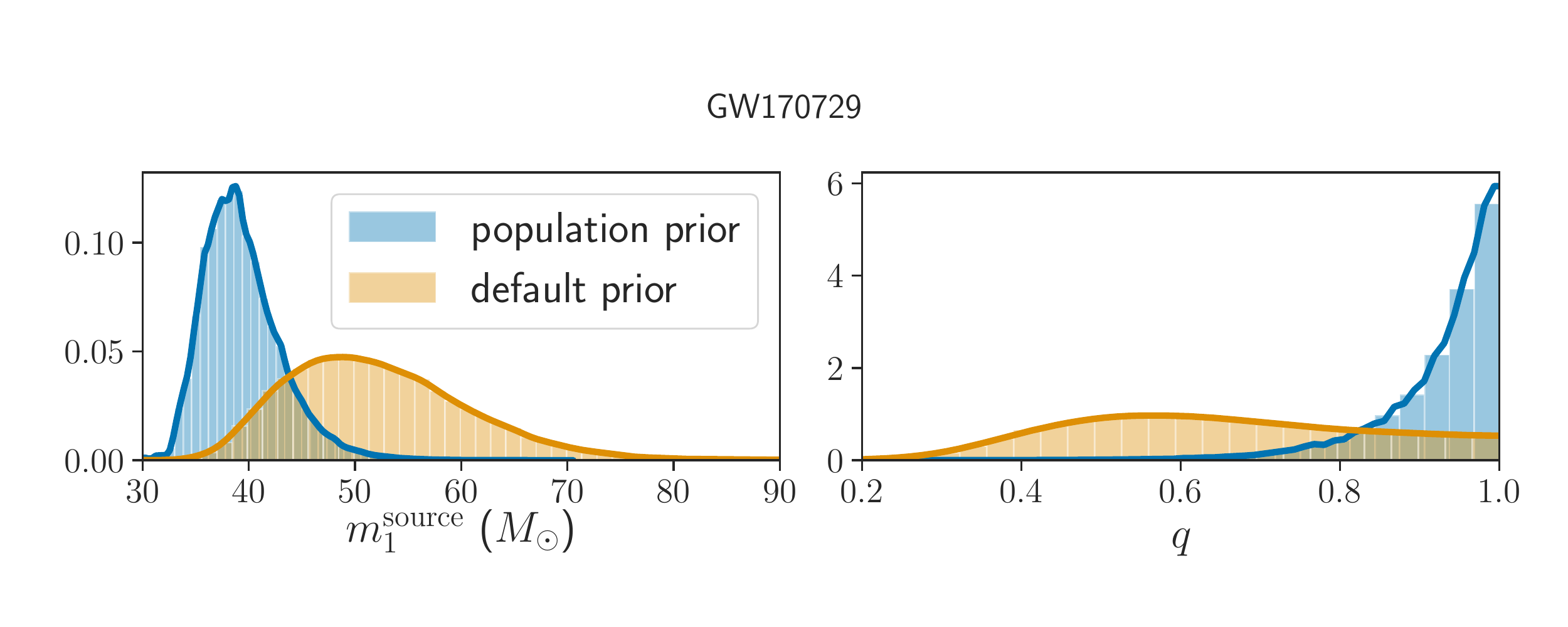}
    \caption{Primary mass (left panel) and mass ratio (right panel) of the BBH event GW170729 under the default (flat in detector-frame masses) prior (orange) \daniel{[although i think i would prefer green]} versus the prior implied by the Model B population analysis of \citet{2018arXiv181112940T}. The population analysis strongly constrains the maximum mass of the population to $\lesssim 50 \ M_\odot$, and favors near-unity mass ratios, which implies that, if we believe the parametric model is a reasonable description of the population, the primary mass and mass ratio of GW170729 are relatively well-constrained. Alternatively, the analysis with the default prior, which ignores the rest of the detected population, ascribes a high probability for an outlier value to the mass.}
    \label{fig:GW170729}
\end{figure*}

In certain cases, the population analysis returns a noticeably different posterior for single-event parameters compared to the posterior returned from PE. For example, for the most massive event of O1 and O2, GW170729, the population analysis implies a much tighter prior on the masses compared to the uninformative priors of PE. The population-informed posteriors on $m_1$ and $q$ for GW170729 under Model B are shown in Figure~\ref{fig:GW170729}. The Model B population analysis implies that the primary mass is $m_1 = \julymassppop$ and the mass ratio is $q = \julyqpop$ compared to $m_1 = \julymasspLI$ and $q = \julyqLI$ under the default uninformative priors.\footnote{\pp{We use the parameter estimation results derived using the {\tt IMRPhenomPv2}~\citep{2016PhRvD..93d4007K} waveform approximant throughout. PE posteriors and population results with the {\tt SEOBNRv3}~\citep{2014PhRvD..89h4006P} approximant are available as well, and there are no significant differences between the two.}}

\section{Evaluating tension between model and data}
\label{sec:PPD}
The above section assumes that the full set of observations can be adequately described by a given model, and it therefore makes sense to impose population priors and recalculate the posteriors on the parameters of each event. However, we often want to explicitly check whether the model fits the data sufficiently and whether the observations are consistent with one another under the model. In this section, we detail various methods of carrying out goodness-of-fit and outlier identification tests, and apply them to the BBH mass distribution fits from~\cite{2018arXiv181112940T}.
\subsection{Definitions and assumptions}
In a hierarchical population analysis, there are three levels at which we can perform a model goodness-of-fit/consistency test. The highest level consists of the population parameters $\Lambda$ and their inferred values from the data. If fitting the population model separately on different subsets of events yields posteriors on the population parameters that are in significant tension with one another, one can conclude that the model does not fit all events as a single population. An example of this test was carried out for Model A of~\citet{2018arXiv181112940T} with a leave-one-out analysis, in which the population hyper-parameters were fit with and without GW170729. Comparing the posteriors on the hyper-parameters with and without GW170729 show that excluding GW170729 from the fit results in statistically consistent posteriors, leading to the conclusion that GW170729 is not a population outlier.

The second level of a hierarchical analysis consists of the (true) event-level parameters $\{\theta_i\}$.
Following~\citet{2018arXiv181112940T}, we define the posterior population distribution as a probability density function (pdf) on the true parameters ${\theta}$ of an (unobserved, and potentially unobservable) system that belongs to the population, given the data we have already observed:
\begin{equation}
\label{eq:ppopd}
    p({\theta} \mid \{ d_i \}) = \int p_\mathrm{pop}({\theta} \mid \Lambda) p(\Lambda \mid \{ d_i \}) \mathrm{d} \Lambda,
\end{equation}
where the posterior $p(\Lambda \mid \{ d_i \})$ is given by marginalizing Eq.~\ref{eq:jointlikelihood} over the event-level parameters $\{ \theta_i \}_{i=1}^{N}$ of the $N$ detected events.
The pdf on the true parameters $\theta$ of a \emph{detection} must take into account selection effects by weighting each $\theta$ by the detectability of the data $\tilde{d}$ that it would produce in our detectors:
\begin{equation}
\label{eq:ppdtrue}
\begin{split}
    p(\theta, \mathrm{det} \mid \{ d_i \}) &=p({\theta} \mid \{ d_i \} ) P_\mathrm{det}(\theta) \\
    &=  p({\theta} \mid \{ d_i \} ) \int P_\mathrm{det}(\tilde{d}) p(\tilde{d} \mid {\theta})  \mathrm{d} {\tilde{d}}.
\end{split}
\end{equation}
Note that the first term outside of the integral above is the posterior population distribution given by Eq.~\ref{eq:ppd}. The term $p(\tilde{d} \mid {\theta})$ takes into account the measurement uncertainty in going from the true parameters of the system ${\theta}$ to the data $\tilde{d}$, and the term $P_\mathrm{det}(\tilde{d})$ accounts for the fact that only some pieces of data are detectable. Throughout, we assume that the detectability of a piece of data, $P_\mathrm{det}(\tilde{d})$, is deterministic, meaning it is always 0 or 1, depending on whether the data pass a (known) detection threshold; this is discussed in more detail below. The term $P_\mathrm{det}(\theta)$ also appears in Eq.~\ref{eq:jointlikelihood}. \pp{Sometimes $VT(\theta)$, the sensitive spacetime volume to a system with parameters $\theta$, appears in place of $P_\mathrm{det}(\theta)$, as these terms are proportional to each other assuming the merger rate is constant in comoving volume and source-frame time.}
Given a collection of events with data $\{ d_i \}$ and a population model, we can calculate the above pdf for the true parameters of detected events (Eq.~\ref{eq:ppdtrue}). Comparing the true parameters $\{\theta_i\}$ of the detected events (as inferred jointly with the population hyperparameters via Eq.~\ref{eq:jointlikelihood}) against the posterior predictive pdf $p(\theta, {\rm det} \mid \{d_i\})$ provides another measurement of the consistency of the population model with the observations.

The third and final level of a hierarchical analysis consists of the data; this is the level on which we will evaluate the population fits for the remainder of this work.
By folding in measurement uncertainty as well as the detection efficiency, we arrive at a probability distribution on the data $\tilde{d}$ from a future detection, rather than on the true parameters. We refer to this as the posterior predictive distribution:
\begin{equation}
\label{eq:ppd}
    p(\tilde{d}, \mathrm{det} \mid \{ d_i \}) = \int P_\mathrm{det} (\tilde{d}) p(\tilde{d} \mid {\theta}) p({\theta} \mid \{ d_i \} ) \mathrm{d} {\theta}.
\end{equation}

In this work, we approximate the GW detection process by assuming that the detection criterion is a fixed threshold on the observed signal-to-noise ratio (SNR) $\rho_\mathrm{obs}$. A detected piece of data ${d}$ from a BBH merger is a timeseries consisting of the GW signal $h(t)$, which, assuming GR, is fully described by the source parameters ${\theta}$ and detector noise $n(t)$. Therefore, we shall identify the data $d$ with the set of observed \pp{(maximum-likelihood)} source parameters $\theta_\mathrm{obs}$, by which we mean the true source parameters offset by some measurement noise: ${d} = \{{m_1}^\mathrm{obs}, {m_2}^\mathrm{obs}, z^\mathrm{obs}, \ldots \} = {\theta}_\mathrm{obs}$.
Once an event is detected, the PE analysis returns a posterior pdf on the source parameters $p({\theta} \mid \tilde{d})$~\citep{2015PhRvD..91d2003V}. In referring to the observed parameters ${\theta}_\mathrm{obs}$ in this work, we mean the maximum likelihood point-values as returned by PE (or, equivalently, the peak of the posterior if the PE prior $\pi({\theta})$ is flat).

The details of how we simulate mock GW datasets with realistic measurement uncertainty in order to compute $P_\mathrm{det}(\tilde{d})$ and $p(\tilde{d} \mid {\theta})$ are found in Appendix~\ref{subsec:appendix1}.
Neglecting spins, each source has true parameters $\theta = \{ m_1, m_2, z, \Theta \}$, where $\Theta$ is a geometric factor encoding a combination of sky position and inclination~\pp{\citep{1993PhRvD..47.2198F}}, such that the amplitude of the GW signal, or signal-to-noise ratio (SNR) $\rho$ is a deterministic function of $m_1$, $m_2$, $z$, and $\Theta$.
Based on the measurement uncertainties described in Appendix~\ref{subsec:appendix1}, the source is then assigned a set of observed parameters $\theta_\mathrm{obs} = \{ m_{1, det}^\mathrm{obs}, m_{2, det}^\mathrm{obs}, \Theta_\mathrm{obs}, \rho_\mathrm{obs} \}$ where $m_{det}$ denotes the detector-frame masses, related to the source-frame masses by a factor of $(1+z)$. These four numbers are in turn used to measure the distance, which is inversely proportional to the observed SNR. Knowing the distance is equivalent to knowing the redshift under a fixed cosmology, and the redshift allows us to deduce the source-frame masses from the detector-frame masses. This is how source-frame masses are inferred for a BBH GW signal under a waveform model.

Recall that when we refer to observed masses in this work, we mean the maximum likelihood values, which do not correspond to the peak of the posteriors shown for event parameters in~\citet{2018arXiv181112907T} because a nonflat prior $\pi(m_1,m_2,z)$ is used. The default prior in the event-level analysis is proportional to the square of the luminosity distance, which places more weight at high $z$ than a flat prior, so the maximum-likelihood source-frame mass values are larger than the maximum a posteriori values.

\pp{We stress that {\em noise fluctuations will generally cause us to estimate erroneously large source-frame masses.}} A notable consequence of the detection threshold is that near-threshold sources are detectable only if a favorable noise fluctuation pulls the observed SNR above the detection threshold. Because the distance measurement correlates with the observed SNR, the systematic shift to larger SNR caused by the detection threshold implies that the observed redshift (as inferred from the distance) tends to be systematically shifted towards smaller values. This in turn implies that the observed source-frame masses, which are inferred by un-redshifting the detector-frame masses, are preferentially shifted towards larger values.

Although extreme noise fluctuations are intrinsically rare, they are statistically guaranteed to affect some fraction of detections. The larger the sample of detections, the larger the fluctuation it contains.

\subsection{Application to the O1+O2 population}
In this subsection, we calculate the posterior predictive mass distribution for the BBH population from O1 and O2 and explore whether the population models provide an adequate fit to the ten BBH observations. We focus on their primary masses, and more specifically, the primary mass of GW170729.

\begin{figure*}
\centering
\includegraphics[width=\textwidth]{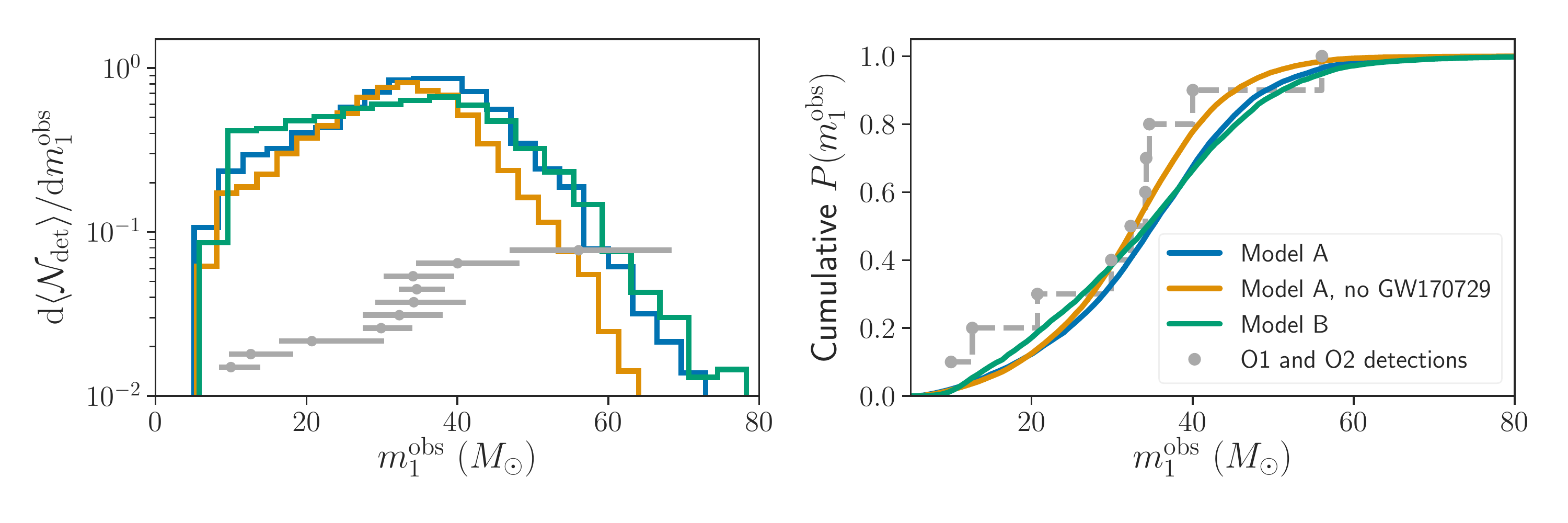}
\caption{\emph{Left panel:} The posterior predictive distribution for the number of detections per \emph{observed} primary mass bin during O1 and O2, based on the fits to Model A and Model B using all ten BBHs (blue and green lines) and the nine BBHs excluding GW170729 (yellow). The gray error bars show the maximum-likelihood points and 90\% credible intervals on the masses of each of the ten BBHs\pp{; the vertical placement of these error bars is arbitrary}. The observed mass is defined as the maximum likelihood estimate of $m_1$, and is predicted according to the synthetic detection and PE process described in the text. \emph{Right panel:} The cumulative posterior predictive distribution, or the probability that an observed mass is less than $m_1^\mathrm{obs}$, inferred from the detections and the given population model, compared to the empirical distribution function from the ten detections in gray (with the gray points denoting the maximum likelihood $m_1$ estimates). The agreement between the observations and each model can be quantified by the distances between the gray points and the colored curve of interest, as calculated in the text. The predictions of the population model match the observations fairly well, and GW170729 does not appear to be an outlier even when excluding it from the calculation of the posterior predictive distribution (yellow curve).}
\label{fig:dNdetdm1obs}
\end{figure*}

The posterior predictive distribution (Eq.~\ref{eq:ppd}) for $m_1^\mathrm{obs}$, given the events from O1 and O2 and the assumed mass Models A and B from~\citep{2018arXiv181112940T}, is shown in Figure~\ref{fig:dNdetdm1obs}. We focus on Models A and B, the simple power-law models, because they have fewer free parameters than Model C, and we wish to check whether these few parameters sufficiently fit the data. We show the posterior predictive distributions as inferred from all ten BBHs, as well as under Model A excluding GW170729. The left panel in Figure~\ref{fig:dNdetdm1obs} shows the expected number of detections per $m_1^\mathrm{obs}$ bin during O1 and O2 according to the model; this is based on the inferred merger rate together with the shape of the mass distribution.

We note that although Model A (B) predicts a sharp cutoff at $\mmax = 41.6^{+9.6}_{-4.3} \ M_\odot$  ($\mmax = 40.8^{+11.8}_{-4.4} \ M_\odot$), and all three models considered in~\cite{2018arXiv181112940T} predict that 99\% of BBHs have $m_1 < 45 \ M_\odot$, Model A (B) predicts that 18\% (20\%) of detected systems have an \emph{observed} primary mass $m_1^\mathrm{obs} > 45 \ M_\odot$. This is because out of the underlying population of BBHs, more massive systems are more likely to be detected, and out of those that are detected, statistical fluctuations can push the observed (maximum likelihood) $m_1^\mathrm{obs}$ to values that are oftentimes significantly larger than the true $m_1$. Recall that among detected sources, statistical fluctuations are more likely to push the observed source-frame masses to larger values than smaller values, because sources near threshold are only detectable if a fluctuation increases their observed SNR, leading to a smaller inferred redshift and a larger inferred source-frame mass. The full likelihood distribution should still have nonzero support at the true value, but for very large statistical fluctuations, the support at the true value may be very small and difficult to resolve.

From Figure~\ref{fig:dNdetdm1obs}, one can visually compare the number and observed masses of the O1 and O2 detections (gray errorbars) to the model predictions; this serves as a posterior predictive check that the model fits the data sensibly. The gray points at the center of the errorbars denote the maximum likelihood $m_1^\mathrm{obs}$ for each event; these points are the values that, according to the model, should be representative draws from the posterior predictive distributions. (The errorbars  denote the 90\% symmetric credible intervals from the full $m_1$ posteriors and are shown only for reference).

The right panel of Figure~\ref{fig:dNdetdm1obs} shows the cumulative, normalized versions of the posterior predictive distributions in the right panel (the colored curves) compared to the empirical distribution function (edf; the gray points). The edf is a cumulative histogram of the maximum likelihood $m_1$, $\hat{m}_{1,i}^\mathrm{obs}$, for each event $i$, defined as:
\begin{equation}
\label{eq:EDF}
\hat{F}_n(m_1^\mathrm{obs}) = \frac{1}{n} \sum_i^n  I[\hat{m}_{1,i}^\mathrm{obs} < m_1^\mathrm{obs}],
\end{equation}
where $n$ is the number of events and $I$ is the indicator function which evaluates to $1$ if its argument is True and zero otherwise.

As seen in the right panel of Figure~\ref{fig:dNdetdm1obs}, the edf appears to follow the posterior predictive cdfs for Models A and B, with perhaps slightly more low-mass and high-mass detections than predicted under the simple power-law models. The relative lack of intermediate-mass detections (and comparable abundance of detections at the low- and high-mass ends of the $\mmin$-$\mmax$ range) can also be seen in the fit to Model C (the tapered power-law with a high-mass Gaussian) of~\cite{2018arXiv181112940T}. Under Model C, the only model that allows for deviations from a pure power-law in $m_1$,~\cite{2018arXiv181112940T} find that the data mildly prefers a merger rate that decreases at intermediate $m_1$ and then rises again at $m_1 \sim 30 \ M_\odot$ (see the top panel of their Figure 2). \cite{2018arXiv181112940T} conclude that this preference for a power-law deviation is not statistically significant because all three models A, B, and C predict distributions of $m_1$ that overlap within the 90\% statistical uncertainties. \cite{2018arXiv181112940T} also report Bayes' factors between all three models and find that a deviation from a power-law is preferred by a factor of $e^{1.92} \approx 7$, although as usual, the values of Bayes' factors are sensitive to the priors chosen for the model hyper-parameters. Our posterior predictive checks for the pure power-law Models A and B do not rely on an explicit comparison to an alternate model, but our conclusions agree with~\cite{2018arXiv181112940T} that the observed $m_1$ distribution is consistent with these power-law models well within the 90\% level.

\pp{More quantitatively,} the distance between the edf and the cumulative probability distributions (cdf) predicted from the models as fit to the data is a measure of ``goodness-of-fit," or how well each model explains the data. This is the basis of the well-known Kolmogorov-Smirnov (KS)~\citep{Kolmogorov,Smirnov} and Anderson-Darling (AD) statistics~\citep{Anderson}. The KS statistic is the maximum distance between the cdf of the model and the edf derived from the data (or two edfs for two different datasets) while the AD statistic is a weighted average of the distance between each point in the sample and the cdf. The KS statistic is mainly sensitive to differences between the centers of the two distributions while the AD statistic is more sensitive to differences in the tails of the distribution~\citep{Stephens}.

Restricting ourselves only to the $m_1$ distribution, we can use the KS and AD statistics to quantify whether the ten observations $\hat{m}_{1,i}^\mathrm{obs}$ are consistent with random draws from the posterior predictive distribution for each population model. If they are inconsistent, the model has failed the posterior predictive check and is a poor fit to the data. We first construct a null distribution, representing the hypothetical distribution of statistics for data that is accurately described by the model. We Monte-Carlo samples from the null distribution by repeatedly drawing sets of ten observations from the posterior predictive distribution of interest and calculating the KS and AD statistic for each set. We then compare the KS and AD statistic for the actual set of ten observations against the null distribution. If the KS and AD statistics for our data is larger than 99\% of cases in the null distribution, we conclude that the probability of the data being described by the model is $<1\%$; in other words, we can reject the particular model with a p-value of 0.01.

Explicitly, we find that the KS statistic between the edf and the Model B posterior predictive cdf lies at $\KSpB$ of the null distribution, while AD statistic lies at $\ADpB$ of its null distribution. For Model A, the KS (AD) statistic is at $\KSpA$ ($\ADpA$) of its null distribution. We also compute these statistics between the edf for all ten events and the Model A posterior predictive cdf as inferred without GW170729. This does not affect the statistics significantly, with the KS statistic corresponding to $\KSpAexc$ of the null distribution and the AD statistic at $\ADpAexc$ in this case, which indicates that the observed primary mass of GW170729 is consistent with the $m_1$ distribution as inferred without it. Counterintuitively, the shift towards smaller KS and AD statistics seems to indicate that the $m_1$ posterior predictive distribution as inferred from only nine events actually provides a slightly better fit to the observed primary masses of all ten events; however, the shift in these statistics is not significant, and illustrates that this is a simplified one-dimensional test that is only meant to check consistency between data and model and not to properly fit a model. \daniel{[I'm very suprised by this]} {\em In summary, Models A, B, and C are all found to be consistent with the observed population.}

Note that the edf is not uniquely defined in two or more dimensions, and Figure~\ref{fig:dNdetdm1obs} shows only how the model fits the observed $m_1$ distribution, ignoring all other parameters of BBH systems including the mass ratio, spin, and redshift.
However, the edf-based test described above can be extended (non-uniquely) to an arbitrary number of dimensions as long as the algorithm for ordering the observations is fixed in advanced. The null distribution can still be calculated according to this fixed ordering algorithm, and, if desired, a p-value can be calculated to evaluate goodness-of-fit.

We now turn to the more specific issue of quantifying whether a particular observation is an outlier with respect to other events that make up the population. For example, in order to investigate the degree of tension between a particularly massive observation, such as GW170729, and the maximum mass inferred from the population analysis, it is useful to consider the posterior predictive distribution of the \emph{maximum} observed primary mass out of $N$ detections, $\mathrm{max}\left(\{m_{1,\mathrm{obs}}^j\}_{j=1}^{N}\right)$, based on the data from the remaining $N-1$ detections, $\{ d_i \}_{i=1}^{N-1}$:
\begin{widetext}
\begin{equation}
\label{eq:ppdmaxm}
    p\left(\left.\mathrm{max}\left(\{m_{1,\mathrm{obs}}^j\}_{j=1}^{N}\right) \right| \{ d_i \}_{i=1}^{N-1}\right) =
    \frac{\mathrm{d}}{\mathrm{d} m} \left. \left( \left[\int_{0}^m p(m_{1,\mathrm{obs}} \left| \{ d_i \}_{i=1}^{N-1}\right.) \mathrm{d} m_{1,\mathrm{obs}}\right]^{N-1}\right) \right|_{m={\mathrm{max}\left( \{ m_{1,\mathrm{obs}}^j \}_{j=1}^{N} \right) }},
\end{equation}
\end{widetext}
where the pdf in the integral is simply the posterior predictive distribution. \pp{Equation~\ref{eq:ppdmaxm} follows from the fact that given $N$ random draws $X_i$ from a fixed pdf $p(x)$ with corresponding cdf $P(X_i < x)$, the maximum draw $Y$ will follow a cdf:
\begin{equation}
P(Y < x) = P(X_i < x)_{\forall i \in [1,N]} = P(X_i < x)^N.
\end{equation}}

\begin{figure}
\includegraphics[width=0.5\textwidth]{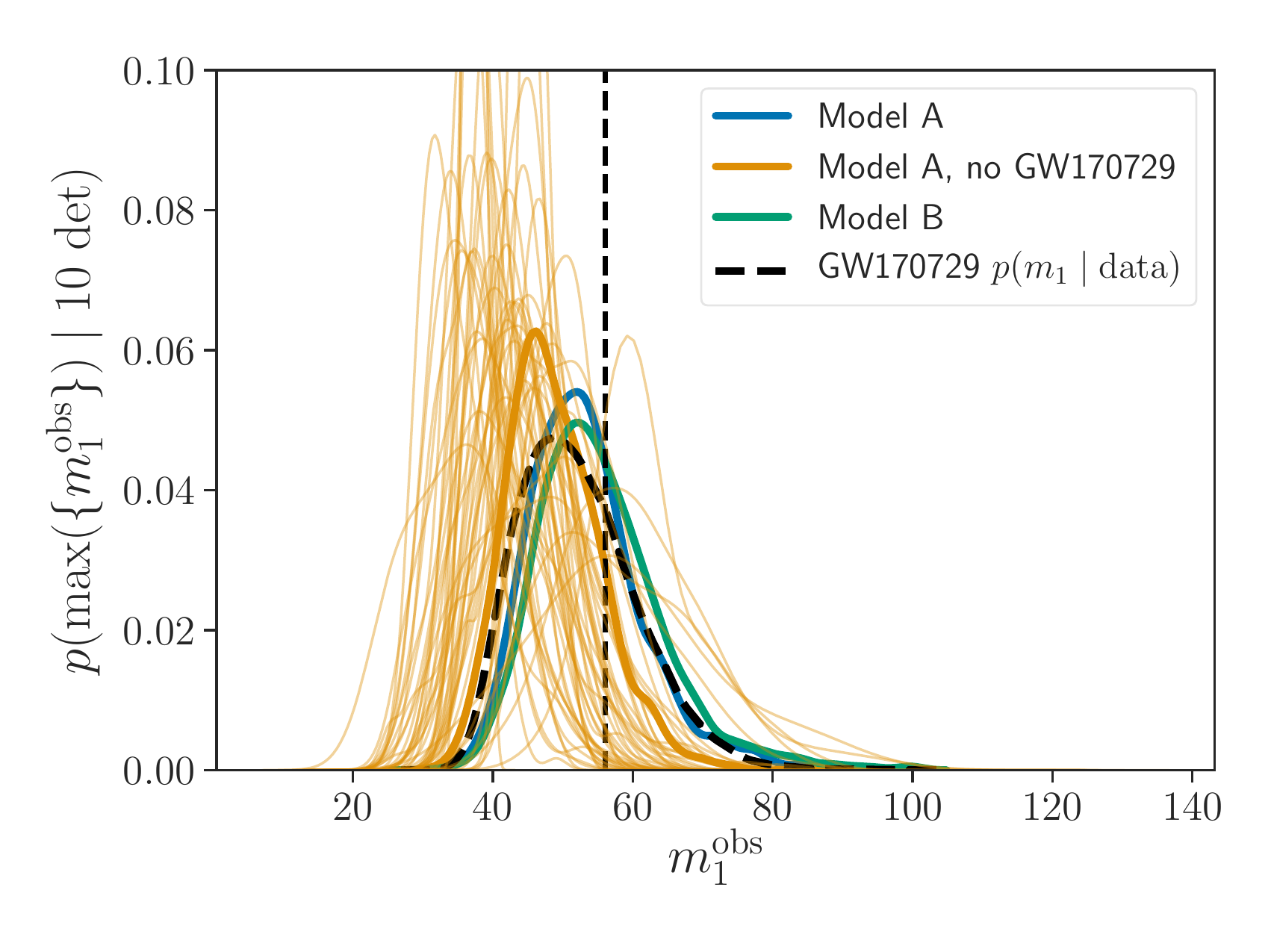}
\caption{Posterior predictive distribution of the maximum \emph{observed} mass out of ten detections as inferred from the detections and the population model of interest (bold, solid colored curves). The observed mass refers to the maximum likelihood $m_1$ value of a detected event as predicted according to a synthetic detection and PE process (see text). The thin orange curves show mock PE posteriors for 50 random events drawn from the bold orange curve, representing 50 examples of posteriors for the most massive $m_1$ that we expect to detect based on the fit to Model A from the nine detections excluding GW170729. For comparison, the posterior for the primary mass of GW170729 is shown (dashed black curve) with the maximum-likelihood value (vertical dashed line). Visually, the GW170729 $m_1$ posterior appears consistent with the thin orange curves. Quantitatively, comparing its maximum-likelihood value to the bold orange curve shows that the primary mass of GW170729 is consistent with the population as inferred from the other nine events at the 86\% level.}
\label{fig:pmaxm1obs_10det}
\end{figure}
The posterior predictive distributions for the maximum observed mass out of ten detections is shown in Figure~\ref{fig:pmaxm1obs_10det}.
The maximum likelihood primary mass $m_1^\mathrm{obs}$ for GW170729 (vertical dashed line) is consistent at the $\julymppdpAexc \%$ level with the posterior predictive distribution on the maximum of ten primary mass observations, as inferred from Model A and the remaining nine observations (orange solid curve). The light orange curves show mock posteriors on $m_1$ for fifty events drawn from the solid orange curve, and we can see that the $m_1$ posterior for GW170729 (dashed black curve) looks like a typical observation. When using GW170729 itself in calculating this posterior predictive distribution, its primary mass is consistent with being the maximum of ten observations at $\julymppdpA \%$ under Model A  (blue curve) and $\julymppdpB \%$ under Model B (green curve).
Based on this analysis, we conclude that there is no evidence for tension between the primary mass of GW170729 and the remaining nine observations under Models A and B, in agreement with the leave-one-out analysis of~\cite{2018arXiv181112940T}.

\begin{figure}
\includegraphics[width=0.5\textwidth]{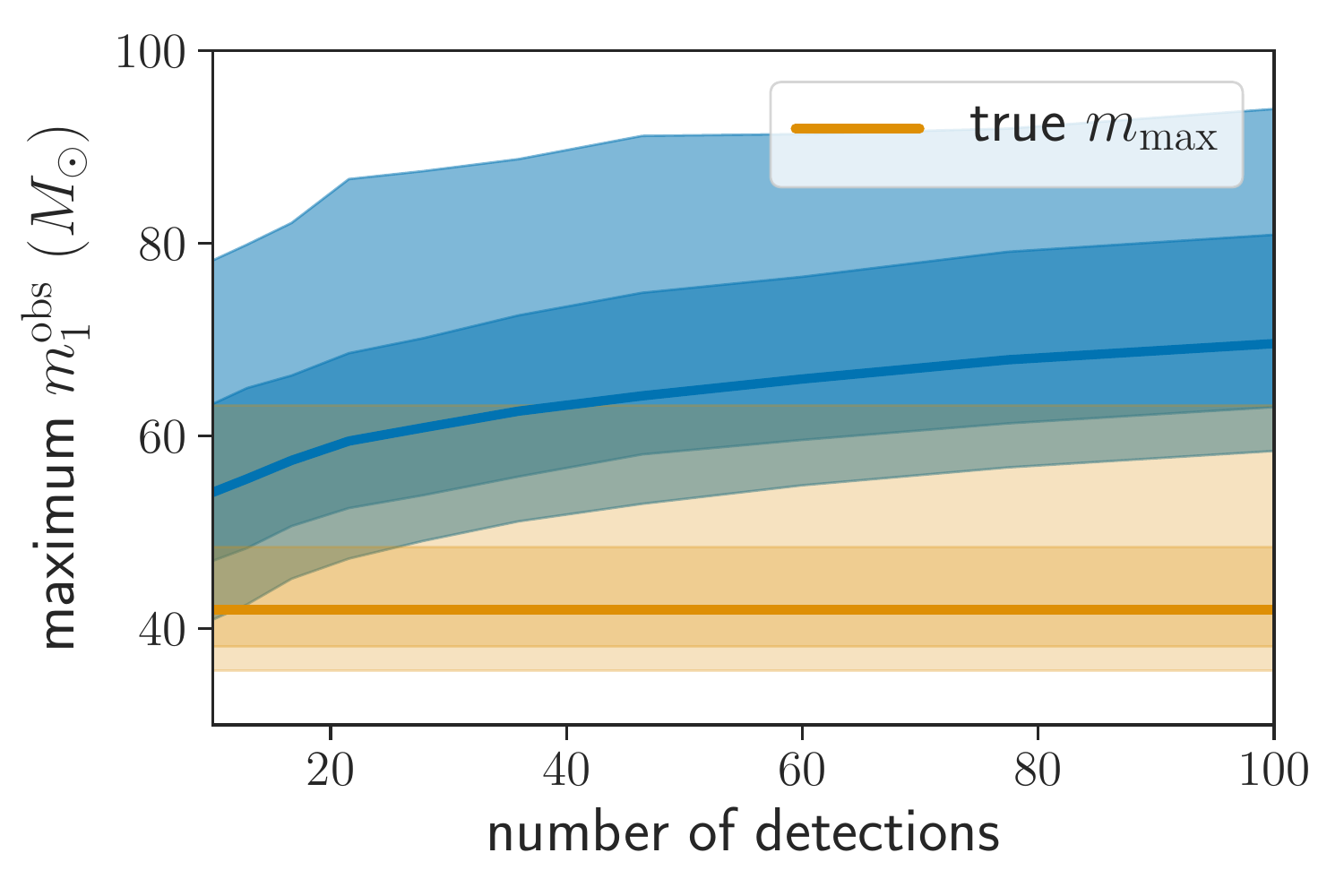}
\caption{Maximum mass we expect to observe as a function of number of detections (blue), as inferred from the Model B fit to the ten O1/O2 detections (which predicts the true $\mmax = 40.8^{+11.8}_{-4.4} \ M_\odot$, shown in orange). The observed mass $m_1^\mathrm{obs}$ represents the maximum likelihood value that would be inferred for the detected system, taking into account measurement uncertainty and selection effects. The solid line denotes the median and dark and light bands denote the 68\% and 95\% credible intervals of the posterior predictive distribution. As the number of detections increases, the largest noise fluctuation in the sample will become more extreme. Furthermore, because of the SNR threshold, these noise fluctuations statistically lead to larger inferred masses.  The blue curves are well fit by $\max m_1^\mathrm{obs} \simeq \max m_1  + \sigma \sqrt{2 \log N}$ with $\sigma$ the observational uncertainty (see text).}
\label{fig:maxm1obs_Ndet.pdf}
\end{figure}

As the sample of BBH detections grow, we expect to see more extreme statistical fluctuations, so that we observe primary masses $m_1^\mathrm{obs}$ that are much higher than the true maximum BH mass. In Figure~\ref{fig:maxm1obs_Ndet.pdf}, we show, as a function of the number of detections, the most massive $m_1^\mathrm{obs}$ that we expect in the sample, based on the fit to Model B with the ten O1/O2 detections. With Gaussian observational uncertainties of width $\sigma$ on $N$ detections with true masses near the mass cutoff the largest observed mass will be $\max m_1^\mathrm{obs} \sim \max m_1 + \sigma \sqrt{2 \log N}$; this scaling is observed empirically in Figure 4.  It would be common that by the time we have 100 detections, even if they are all described by Model B with a sharp cutoff at $\mmax = 40.8^{+11.8}_{-4.4} \ M_\odot$, that we will observe a system with $m_1^\mathrm{obs} \sim 70 \, M_\odot$.

\section{Non-parametric Constraints on the Rate of Mass-Gap Mergers}
\label{sec:binned}

\pp{To explore the rate of high-mass mergers in a less parametric way,} we follow the binned-histogram analyses of~\citet{2014ApJ...795...64F} and~\cite{2017MNRAS.465.3254M}. We model the rate of BBH mergers on the two-dimensional $m_1$--$m_2$ plane, $\mathcal{R}(m_1, m_2) = \frac{dN}{dm_1 dm_2 dV_c dt}$ as piecewise constant in $9\times9$ logarithmically-spaced mass bins between 3 and 150~$M_\odot$. The height of each mass bin $\mathcal{R}_{ij}$ represents the merger rate in that bin, and, as in the above sections, we assume the merger rate does not evolve with redshift. We take the prior on the bin heights to be a squared-exponential Gaussian process (GP), where the relative means $\mu_{ij}$ of the bin heights follow a fixed shape $p(m_{1,i}, m_{2,j})$ and the length scales in $\log(m_1)$ and $\log(m_2)$ are fit from the data \pp{(see Appendix~\ref{subsec:appendix2} for more details)}. The point $(\log m_{1,i}, \log m_{2,j})$ denotes the center of the $ij$th bin. We consider two different shapes for the mean merger rate per bin in the GP prior: a  ``power-law" shape prior and ``flat-in-log" shape prior. These are motivated by the two fixed-parameter models that the LVC used to calculate BBH merger rates in O1 and O2~\citep{2016PhRvX...6d1015A, 2018arXiv181112907T}.
In the power-law shape prior, we have:
\begin{equation}
p(m_1, m_2) \propto \frac{m_{1}^{-2.35}}{m_{1} - M_\mathrm{min}},
\end{equation}
or, for the logarithmically-spaced bins,
\begin{equation}
\mu_{ij} = p(m_{1,i}, m_{2,j}) \frac{\mathrm{d}(m_1m_2)}{\mathrm{d}(\log{m_1}\log{m_2})} \propto \frac{m_{1,i}^{-1.35} m_{2,j}}{m_{1,i} - M_\mathrm{min}},
\end{equation}
where $\mmin = 3 \ M_\odot$.
The flat-in-log shape prior is simply:
\begin{equation}
\mu_{ij} \propto 1.
\end{equation}
The joint posterior for the GP prior parameters $\Gamma$, the component masses for each event $\{m_1,m_2\}$, and the merger rate $\mathcal{R}_{ij}$ in bin $ij$ is the inhomogeneous Poission process:
\begin{multline}
  \label{eq:posterior}
  p\left(\mathcal{R}_{ij}, \Gamma, \{m_1,m_2\} \mid d \right) \\ \propto
  \left[ \prod_{k=1}^{N_{\rm events}} p\left( d^{(k)} \mid m_1^{(k)}, m_2^{(k)}
    \right) \mathcal{R}(m_1^{(i)}, m_2^{(i)}) \right]
  \\ \times \exp\left[ -\sum_{ij} \lambda_{ij} \right] p\left( \mathcal{R}_{ij}
    | \Gamma \right) p\left( \Gamma \right),
\end{multline}
where $p\left( d^{(k)} \mid m_1^{(k)}, m_2^{(k)} \right)$ denotes the likelihood of the data for event $k$ given component masses, marginalized over all other parameters of the system, $p\left( \mathcal{R}_{ij} | \Gamma \right)$ represents the GP prior on the bin heights and $\lambda_{ij}$ denotes the expected number of detections in bin ${ij}$, folding in selection effects. See the appendix for more details.

\begin{figure}
\includegraphics[width=0.5\textwidth]{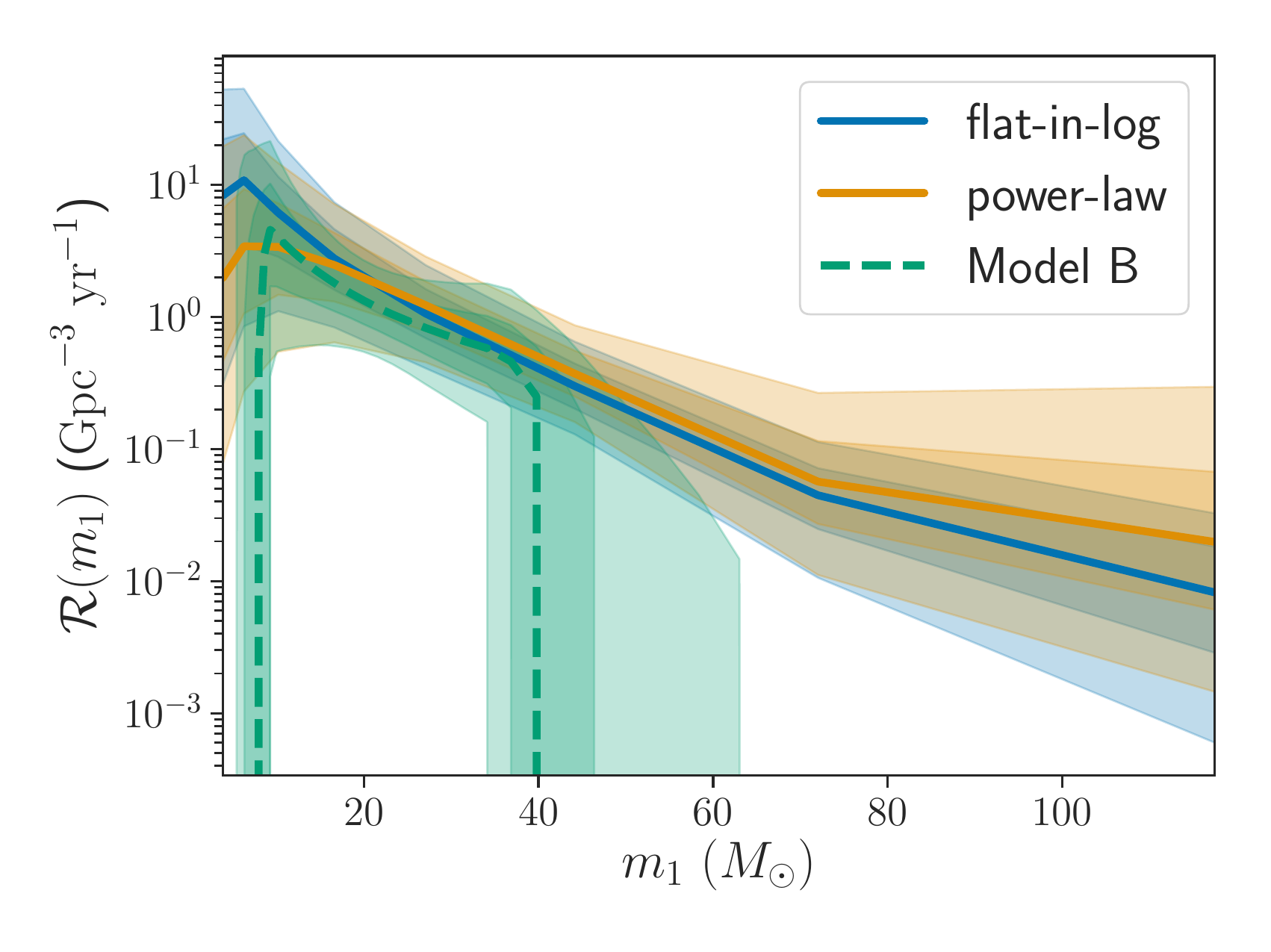}
\caption{The non-parametric constraints on the merger rate distribution for $m_1$ from the binned histogram model under the two shape priors (flat-in-log prior in blue and power-law prior in orange). The solid lines denote the median rate and the dark and light shaded bands denote 68\% and 95\% symmetric credible intervals. In green we show the merger rate as a function of $m_1$ for the parametric mass distribution Model B; this model includes a maximum mass cutoff as one of the parameters. In the low-mass region $m_1 \lesssim 45 \ M_\odot$, the non-parametric model under both priors agrees with the parametric model. Because of the lack of detections at high masses, the parametric Model B infers a tight constraint on $\mmax$ and the merger rate falls to zero, while the nonparametric model attempts to extrapolate smoothly to high masses under strong influence of the prior.}
\label{fig:binned_pm1}
\end{figure}

Figure~\ref{fig:binned_pm1} shows the merger rate $\mathcal{R}(m_1)$ as inferred under the two different priors, where:
\begin{equation}
\mathcal{R}(m_1) = \int_3^{m_1} \mathcal{R}(m_1, m_2) dm_2.
\end{equation}
For comparison we also show the results from the parametrized Model B fit.
We note that in the range $m_1 \lesssim 45 \ M_\odot$, the inferred merger rates $\mathcal{R}(m_1)$ agree between all three models: the two shape priors in the nonparametric model as well as the parametric power-law model. (The difference for $m_1 < 5 \ M_\odot$ is due to the prior on $\mmin > 5 \ M_\odot$ for Model B, while the lowest mass bin in the nonparametric model starts at $3 \ M_\odot$.) Beyond $\sim 45 \ M_\odot$, the merger rate inferred under Model B drops sharply due to the $\mmax$ parameter, whose inferred value closely follows the mass of the most massive observed system~\citep{2017ApJ...851L..25F}. The binned model, on the other, does not have a $\mmax$ parameter that lets the rate fall to zero, and instead has a prior that strongly favors smooth variations of the merger rate from mass bin to mass bin. In the mass bins with $m_1 \gtrsim 45 \ M_\odot$, where there are no detections, the posterior on the merger rate smoothly transitions to following the prior on the bin heights. This is visible in Figure~\ref{fig:binned_pm1} as the merger rate $\mathcal{R}(m_1)$ inferred under the two different priors (blue and orange bands) start to diverge from one another at $m_1 \gtrsim 45 \ M_\odot$, and the uncertainty for each one grows as well. This is a consequence of the GP smoothing prior. After enough detections, if the absence of high-mass events continues, the likelihood will overcome the smoothing prior and the posterior will reveal a sharp drop-off in the merger rate in the binned analysis, independently of the prior on the bin heights. This was demonstrated by~\citet{2017MNRAS.465.3254M} with simulated data in the context of the putative low mass gap between the binary neutron star (BNS) and BBH population. With only ten detections, the binned model provides a conservative upper limit on the rate of mergers with $45 \ M_\odot < m_1 < 150  \ M_\odot$ under the prior that the merger rate should not vary sharply between neighboring mass bins.

Under the flat-in-log prior, we infer the merger rate in the mass range $45 \ M_\odot < m_1 < 150  \ M_\odot$ to be $3.02^{+12.97}_{-2.28}$ Gpc$^{-3}$ yr$^{-1}$ (90\% equal-tailed credible interval), or
$1.79^{+2.30}_{-1.23}$ Gpc$^{-3}$ yr$^{-1}$ under the power-law prior.
We infer the total merger rate over the $3 \ M_\odot < m_1 < 150 \ M_\odot$ range to be $42.50^{+68.12}_{-24.89}$ Gpc$^{-3}$ yr$^{-1}$ with the flat-in-log shape prior or $65.58^{+102.34}_{-41.82}$ Gpc$^{-3}$ yr$^{-1}$ with the power-law shape prior, implying that the rate of mergers with $m_1 > 45 M_\odot$ makes up $7.6\%^{+23.8}_{-6.0}$ (flat-in-log prior) or $2.8\%^{+5.4}_{-2.0}$ (power-law prior) of the total merger rate. This is to be compared to the parametric models of~\cite{2018arXiv181112940T}, which all predict that less than 1\% of mergers have $m_1 > 45 \ M_\odot$. \pp{Unlike the parametric models with a maximum mass parameter, the binned-histogram model does not allow the high-mass merger rate to drop all the way to zero.} The advantage of the nonparametric constraints is that, if there were a secondary population of BBH mergers that does not respect the maximum mass feature of the parametrized mass models (consisting, for example, of second-generation mergers that occupy the pair-instability mass gap), we still expect their merger rate to respect these nonparametric limits.

\section{Conclusion}
\label{sec:conclusion}
Focusing on the BBH mass distribution as inferred by~\cite{2018arXiv181112940T}, we have explored how individual events fit into a population analysis, especially in the presence of measurement uncertainty and selection effects. We have presented simple posterior predictive/goodness-of-fit checks to show consistency between the O1/O2 events and the power-law mass distribution models of~\cite{2018arXiv181112940T}. In particular, GW170729, the most massive event of O1/O2, is not an outlier with respect to even the simplest power-law model with a sharp high-mass
cutoff. When folding in the full information about the population, the primary mass of GW170719 is inferred to be $m_1 = \julymassppop$ as compared to the inferred value under the uninformative PE priors, $m_1 = \julymasspLI$. Although the simple parametrized models provide adequate fits to the BBH detections so far, we have presented nonparametric fits to the mass distribution based on the GP-regularized binned-histogram model of~\cite{2017MNRAS.465.3254M}. Under this model, we place conservative upper limits on the rate of mergers with $45 \ M_\odot < m_1 < 150 \ M_\odot$, and find that these high-mass mergers make up at most $7.6\%^{+23.8}_{-6.0}$ of the total merger rate in the range $3 \ M_\odot < m_1 < 150 \ M_\odot$ for a flat-in-log shape prior on the mass distribution, or $2.8\%^{+5.4}_{-2.0}$ of the total merger rate for a power-law shape prior.

\acknowledgments
We thank Tom Callister for his helpful comments on the manuscript. MF was supported by the NSF Graduate Research Fellowship Program under grant DGE-1746045. MF and DEH were supported by NSF grant PHY-1708081. They were also supported by the Kavli Institute for Cosmological Physics at the University of Chicago through NSF grant PHY-1125897 and an endowment from the Kavli Foundation. Part of this work was completed at the Kavli Institute for Theoretical Physics, supported by NSF grant PHY-1748958. DEH also gratefully acknowledges support from the Marion and Stuart Rice Award.

\appendix{}
\label{sec:appendix}
\section{Mock observations}
\label{subsec:appendix1}
This section explains in greater detail how we calculate selection effects and simulate measurement uncertainty for mock observations.
For a BBH with true parameters $\theta$, we follow the simple prescription of~\citet{2018ApJ...863L..41F} and~\citet{2019arXiv190809084F} to assign realistic measurement uncertainty and compute $\theta_\mathrm{obs}$. Given the BBH's source-frame masses, spins, and redshift together with a PSD describing the noise of the detector, we can calculate the optimal SNR $\rho_\mathrm{opt}$ of the source, which is the SNR that it would have if it were optimally oriented face-on and directly overhead of the detector~\citep{2017arXiv170908079C}. We assume a fixed cosmology described by the best-fit Planck 2015 parameters~\citep{2016A&A...594A..13P} to interchange between redshift and luminosity distance~\citep{astropy}.  We fix the PSD to the ``aLIGO Early High Sensitivity" noise curve from~\citet{2018LRR....21....3A}, which is representative for O1 and O2. We also fix BBH spins to zero in this calculation, since they have a negligible effect on the SNR calculation for population studies~\citep{2018arXiv181112940T}. An isotropic distribution of sky positions and inclinations relative to a detector yields a distribution of true SNRs $\rho$ in that detector; this distribution can be summarized by the angular projection term $ 0 \leq \Theta = \frac{\rho}{\rho_\mathrm{opt}} \leq 1$. The angular term $\Theta$ has a known distribution~\citep{1993PhRvD..47.2198F}. Therefore, for a BBH with intrinsic parameters $\{ m_1, m_2, z \}$, we assign a single extrinsic parameter $\Theta$ drawn from this distribution. These four parameters together correspond to a true SNR $\rho$ for the source.

Given $\{m_1, m_2, z, \Theta, \rho \}$ for a source, we assign measurement uncertainty as follows. We first draw an observed SNR $\rho_\mathrm{obs}$ from a normal distribution centered at the true SNR $\rho$:
\begin{equation}
\label{eq:rho}
\rho_\mathrm{obs} = \rho + N(0,1)
\end{equation}
where $N(\mu,\sigma)$ is a random number drawn from a normal distribution with mean $\mu$ and standard deviation $\sigma$.
We assume that sources are only detected if they pass an SNR threshold (in a single detector) $\rho_\mathrm{obs} > 8$; this would be identical to the semi-analytic selection effects calculation of~\citep{2016ApJ...833L...1A,2016PhRvX...6d1015A,2018arXiv181112940T} were it not for the inclusion of the noise term $N(0,1)$.
To best approximate the mass measurement, we work with the detector-frame (redshifted) chirp mass:
\begin{equation}
\mathcal{M}_z = (1+z)\frac{(m_1m_2)^{3/5}}{(m_1+m_2)^{1/5}},
\end{equation}
and symmetric mass ratio:
\begin{equation}
\eta = \frac{m_1m_2}{(m_1+m_2)^2}.
\end{equation}
The detector-frame chirp mass drives the leading-order GW frequency evolution during the inspiral and is thus the best-measured mass parameter for stellar-mass compact binary sources.
We assume that the uncertainties on the measured parameters scale inversely with $\rho_\mathrm{obs}$, so that:
\begin{align}
\label{eq:obs}
\log \mathcal{M}_z^\mathrm{obs} &= \log \left[\mathcal{M}_z + N(0,\sigma_\mathcal{M}/\rho_\mathrm{obs}) \right],\\
\eta_\mathrm{obs} &= N^{T[0,0.25]}(\eta,\sigma_\eta/\rho_\mathrm{obs}), \\
\Theta_\mathrm{obs} &= N^{T[0,1]}(\Theta,\sigma_\Theta/\rho_\mathrm{obs}).
\end{align}
In the above expressions, $N^{T[a,b]}(\mu, \sigma)$ denotes a random number drawn from the truncated normal distribution. From $\mathcal{M}_z^\mathrm{obs}$ and $\eta_\mathrm{obs}$, we calculate the detector-frame component masses $m_{1,z}^\mathrm{obs}$ and $m_{2,z}^\mathrm{obs}$:
\begin{equation}
m^\mathrm{obs}_{1/2,z} = \frac{M\pm \sqrt{M^2-4\eta M^2}}{2}
\end{equation}
for $M = \mathcal{M}_z^\mathrm{obs}/\eta_\mathrm{obs}^{3/5}$.
The observed redshift in inferred directly from the remaining parameters, via the observed luminosity distance $d_\mathrm{obs}$:
\begin{equation}
\label{eq:dobs}
d_\mathrm{obs} = \frac{d_{\rm fid} \rho_\mathrm{opt}((1+z)m_1^\mathrm{obs},(1+z)m_2^\mathrm{obs},d_{\rm fid})\Theta_\mathrm{obs}}{\rho_\mathrm{obs}},
\end{equation}
where $d_{\rm fid}$ is an arbitrary fiducial luminosity distance (for given detector-frame masses, the SNR of a source scales inversely with its luminosity distance).
The observed redshift $z_\mathrm{obs}$ is derived from $d_\mathrm{obs}$ by the cosmological redshift-luminosity distance relation, and once this is known, we infer the source-frame masses:
\begin{equation}
m_{1/2}^\mathrm{obs} = \frac{m_{1/2,z}^\mathrm{obs}}{1+z_\mathrm{obs}}.
\end{equation}

The observed values $m_1^\mathrm{obs}, m_2^\mathrm{obs}, z_\mathrm{obs}$ denote the maximum likelihood values of the parameters as extracted from the GW signal. To simulate full posterior distributions on these parameters, we use Eq.~\ref{eq:rho}-\ref{eq:dobs} as the likelihood for the observed parameters given true values $\rho$, $\mathcal{M}_z$, $\eta$, and $\Theta$. We take samples on $\rho$, $\mathcal{M}_z$, $\eta$, and $\Theta$ from this likelihood and convert the samples to the space $m_{1,z}, m_{2,z}, d_L, \Theta$, on which we wish to set a prior:
\begin{equation}
\label{eq:PEprior}
\pi_{\rm PE}(m_{1,z}, m_{2,z}, d_L, \Theta) \propto p(\Theta) d_L^2.
\end{equation}
This matches the default PE prior used by LIGO/Virgo in the O1 and O2 event analysis~\citep{2018arXiv181112907T}, where $p(\Theta)$ is the true distribution from which the $\Theta$ values are drawn, representing an isotropic distribution on the sky and source inclination.
The change of variables from $\rho, \mathcal{M}_z, \eta, \Theta$ to $m_{1,z}, m_{2,z}, d_L, \Theta$ means we also have to divide out by the induced prior, given by the Jacobian:
\begin{equation}
\label{eq:jac}
\left|\frac{\mathrm{d}\rho}{\mathrm{d}d_L}\frac{\mathrm{d}(\mathcal{M}_z,\eta)}{\mathrm{d}(m_{1,z},m_{2,z})}\right| = \frac{\Theta \rho_\mathrm{opt}(m_{1,z}, m_{2,z}, d_\mathrm{fid}) d_\mathrm{fid}}{d_L^2} \frac{(m_{1,z}-m_{2,z})\eta(m_{1,z},m_{2,z})^{3/5}}{(m_{1,z}+m_{2,z})^2}
\end{equation}
Once we reweight the $m_{1,z}, m_{2,z}, d_L, \Theta$ samples by Eq.~\ref{eq:PEprior} divided by Eq.~\ref{eq:jac}, we can get posterior samples for the source-frame parameters by converting to $m_1, m_2, z, \Theta$ space.

We tune the $\sigma$ parameters above to match the measurement uncertainties on masses and redshifts found by~\citet{2017PhRvD..95f4053V} when simulating full PE on injected signals.
\citet{2017PhRvD..95f4053V} found that for BBHs detected by Advanced LIGO/ Virgo at design sensitivity, the relative uncertainty (at the 90\% credible interval) on the detector-frame masses is typically $\sim 40\%$ and the relative uncertainty on redshift is typically $\sim 50\%$. However, for the majority of O1 and O2, only the two (co-aligned) LIGO detectors were operational, implying a reduced ability for the network to constrain the polarization of a source and break the distance-inclination degeneracy, and worsened redshift constraints.
We find that for the O1 and O2 events, a more typical relative redshift uncertainty is 70\% (for a  90\% credible interval relative to the median value). We find that choosing $\sigma_\mathcal{M} = 0.08 \rho_{\rm thresh}$, $\sigma_\eta = 0.022 \rho_{\rm thresh}$ and $\sigma_\Theta = 0.21 \rho_{\rm thresh}$ yields measurement uncertainties that match the widths of the O1 and O2 credible intervals and the expectations from~\citet{2017PhRvD..95f4053V}.
We use $\rho_{\rm thresh} = 8$ throughout, as discussed above. The measurement uncertainty on $\Theta$ controls the measurement uncertainty on $z$ according to Eq.~\ref{eq:dobs}. For simulating events for a 3-detector network at design sensitivity, we would use $\sigma_\Theta = 0.15$ to reflect the improved distance constraints of a 3-detector network with relative uncertainties of 50\% rather than 70\%.
We note that the measurement uncertainty on the source-frame masses is a combination of the detector-frame mass uncertainty and the absolute redshift uncertainty, which is largest for sources at high redshift. Therefore, our predictions for distributions of $m_1^\mathrm{obs}$ based on distributions of $m_1$ are sensitive to assumptions regarding the underlying redshift distribution and the network sensitivity, which together determine the detected redshifts of the sources. The uncertainty in these assumptions is subdominant to uncertainties in the population model.

\section{Binned histogram likelihood}
\label{subsec:appendix2}
In this appendix, we provide additional analysis details regarding the binned-histogram fit to the mass distribution (Section~\ref{sec:binned}). The total posterior for the rate $\mathcal{R}_{ij}$ in each mass bin, the parameters $\Gamma$ governing the Gaussian process prior on the bin heights, and the true masses of the detected events is given by Eq.~\ref{eq:posterior}. Below we explain the various terms that appear in this equation.
The GP prior on the (log) of the bin heights takes the form:
\begin{equation}
p(\mathcal{\log R}_{ij} \mid \Gamma) = N(\mathcal{\log R}_{ij}  \mid \log(\mu+\mu_{ij}), \Sigma).
\end{equation}
\pp{where $N(x \mid \hat{\mu},\hat{\Sigma})$ denotes the multivariate normal probability distribution function on $x$} with mean $\hat{\mu}$ and covariance $\hat{\Sigma}$, $\mu$ is an overall scale factor, $\mu_{ij}$ is fixed to either the power-law or flat-in-log shape discussed in Section~\ref{sec:binned}, and $\Sigma$ is a covariance matrix.
We use a squared exponential kernel for the covariance matrix, and parameterize $\Sigma$ as:
\begin{equation}
\label{eq:Sigma}
\Sigma_{ijkl} = (1-f) \sigma^2 \exp\left[ -\frac{1}{2\tau^2}\Delta^2 \pp{\log} m_{ijkl}\right] + f \sigma^2 \delta_{ijkl}.
\end{equation}
For numerical stability, the covariance $\Sigma$ includes some fraction $f \ll 1$ that is white and uncorrelated; the precise value of $f$ does not affect the results and is fixed to $f=0.01$ throughout the analysis. The square of the Euclidean distance between the centers of the bins $(\log m_{1,i}, \log m_{2,j})$ and $(\log m_{1,k}, \log m_{2,l})$ is denoted  $\Delta^2 \pp{\log} m_{ijkl}$. At zero separation, the variance is $\sigma^2$.  The correlation length scale is set by the parameter $\tau$.
Lastly, $\delta_{ijkl}$ is the Kroniker delta function.

In summary, the GP parameter set $\Gamma$ consists of the free parameters $\mu$, $\sigma$, and $\tau$, and the fixed parameters $\mu_{ij}$ and $f$. For the prior on these hyper-parameters (written $p(\Gamma)$ in Eq.~\ref{eq:posterior}), we take a broad Gaussian prior for $\mu$ with mean 0 and standard deviation 10 and a half-Gaussian prior on $\sigma>0$ with mean 0 and standard deviation 1.

We take a Gaussian prior for $\log \tau$.  The mean and standard deviation of
the Gaussian are chosen to place the width of the smallest mass bin 2-$\sigma$
below the mean and the width of the mass range considered ($3 \, M_\odot$ to
$150 \, M_\odot$) 2-$\sigma$ above the mean.  Thus the prior constrains the
correlation length of the rate in log-mass to be typically longer than one bin,
but shorter than the entire mass range. The correlation length $\tau$ is poorly
constrained by the data with only ten detections, and the recovered posterior is
very similar to the prior. We stress that in the limit of a large number of
detections, the likelihood will dominate the GP prior and the posterior on the
bin heights will become independent of these prior choices.

To evaluate the posterior of Eq.~\ref{eq:posterior} we also need the expected number of detections $\lambda_{ij}$ in bin $ij$. To do this we evaluate the sensitive spacetime volume in the $ij$-th bin, $\langle VT \rangle_{ij}$, so that:
\begin{equation}
\lambda_{ij} = \mathcal{R}_{ij} \langle VT \rangle_{ij},
\end{equation}
The sensitive volume was introduced in Section~\ref{sec:PPD}, and is calculated from the detection probability $P_\mathrm{det}$ by~\citep{2018arXiv181112940T}:
\begin{equation}
VT(m_1,m_2) = T_\mathrm{obs} \int P_\mathrm{det}(m_1, m_2, z)\frac{1}{1+z}\frac{\mathrm{d}V_c}{\mathrm{d}z} \mathrm{d}z,
\end{equation}
where $T_\mathrm{obs}$ is the observing time. This assumes that the merger rate is constant (non-evolving with redshift) in comoving volume and source-frame time, and that the detector sensitivity is constant in time.
The average sensitive volume in bin $ij$ is:
\begin{equation}
\langle VT \rangle_{ij} = \int_{m_{i-1/2}}^{m_{i+1/2}} \int_{m_{j-1/2}}^{m_{j+1/2}} VT(m_1,m_2) \frac{\mathrm{d}m_1}{m_1} \frac{ \mathrm{d}m_2}{m_2},
\end{equation}
where the integral boundaries are the edges of the bin.

Consistent with the other analyses in this work, we calculate $P_\mathrm{det}(m_1,m_2,z)$ under the assumption of a single-detector SNR threshold. \cite{2018arXiv181112940T} found this approximation to $VT(m_1,m_2)$ to overestimate the sensitive volume by a factor of $\sim 1.6$ relative to the volume calculated by injections into the detection pipeline. This factor of $\sim 1.6$ between the semi-analytic sensitive volume and the injection-estimated volume is fairly constant across the mass space. We therefore divide our value of $\langle VT\rangle_{ij}$ by 1.6 to match the results of injections. We further assume that the total $\langle VT\rangle$ has a (Gaussian) 1-$\sigma$ error of 18\% (arising from an amplitude calibration error of $\sim6$\%) and marginalize over this uncertainty, split evenly between the mass bins~\citep{2016PhRvX...6d1015A}.

\bibliographystyle{aasjournal}
\bibliography{references}

\begin{thebibliography}{}
\expandafter\ifx\csname natexlab\endcsname\relax\def\natexlab#1{#1}\fi
\providecommand{\url}[1]{\href{#1}{#1}}

\bibitem[{Aasi {et~al.}(2015)}]{TheLIGOScientific:2014jea}
Aasi, J., {et~al.} 2015, Class. Quant. Grav., 32, 074001

\bibitem[{{Abbott} {et~al.}(2018){Abbott}, {Abbott}, {Abbott}, {Abernathy},
  {Acernese}, {Ackley}, \& {Adams}}]{2018LRR....21....3A}
{Abbott}, B.~P., {Abbott}, R., {Abbott}, T.~D., {et~al.} 2018, Living Reviews
  in Relativity, 21, 3

\bibitem[{{Abbott} {et~al.}(2019{\natexlab{a}}){Abbott}, {Abbott}, {Abbott},
  {Abraham}, {Acernese}, {Ackley}, {Adams}, \&
  {Adhikari}}]{2018arXiv181112907T}
---. 2019{\natexlab{a}}, Phys. Rev. X, 9, 031040

\bibitem[{{Abbott} {et~al.}(2019{\natexlab{b}}){Abbott}, {Abbott}, {Abbott},
  {Abraham}, {Acernese}, {Ackley}, {Adams}, \&
  {Adhikari}}]{2018arXiv181112940T}
---. 2019{\natexlab{b}}, \apj, 882, L24

\bibitem[{{Abbott} {et~al.}(2016{\natexlab{a}}){Abbott}, {Abbott}, {Abbott},
  {Abernathy}, {Acernese}, {Ackley}, {Adams}, {Adams}, {Addesso}, {Adhikari},
  {Adya}, {Affeldt}, {Agathos}, {Agatsuma}, {Aggarwal}, {Aguiar}, {Aiello},
  {Ain}, {Ajith}, {Allen}, {Allocca}, {Altin}, {Anderson}, {Anderson}, {Arai},
  {Araya}, {Arceneaux}, {Areeda}, {Arnaud}, {Arun}, {Ascenzi}, {Ashton}, {Ast},
  {Aston}, {Astone}, {Aufmuth}, {Aulbert}, {Babak}, {Bacon}, {Bader}, {Baker},
  {Baldaccini}, {Ballardin}, {Ballmer}, {Barayoga}, {Barclay}, {Barish},
  {Barker}, {Barone}, {Barr}, {Barsotti}, {Barsuglia}, {Barta}, {Bartlett},
  {Bartos}, {Bassiri}, {Basti}, {Batch}, {Baune}, {Bavigadda}, {Bazzan},
  {Bejger}, {Bell}, {Berger}, {Bergmann}, {Berry}, {Bersanetti}, {Bertolini},
  {Betzwieser}, {Bhagwat}, {Bhandare}, {Bilenko}, {Billingsley}, {Birch},
  {Birney}, {Birnholtz}, {Biscans}, {Bisht}, {Bitossi}, {Biwer}, {Bizouard},
  {Blackburn}, {Blair}, {Blair}, {Blair}, {Bloemen}, {Bock}, {Boer}, {Bogaert},
  {Bogan}, {Bohe}, {Bond}, {Bondu}, {Bonnand}, {Boom}, {Bork}, {Boschi},
  {Bose}, {Bouffanais}, {Bozzi}, {Bradaschia}, {Brady}, {Braginsky},
  {Branchesi}, {Brau}, {Briant}, {Brillet}, {Brinkmann}, {Brisson}, {Brockill},
  {Broida}, {Brooks}, {Brown}, {Brown}, {Brown}, {Brunett}, {Buchanan},
  {Buikema}, {Bulik}, {Bulten}, {Buonanno}, {Buskulic}, {Buy}, {Byer},
  {Cabero}, {Cadonati}, {Cagnoli}, {Cahillane}, {Calder{\'o}n Bustillo},
  {Callister}, {Calloni}, {Camp}, {Cannon}, {Cao}, {Capano}, {Capocasa},
  {Carbognani}, {Caride}, {Casanueva Diaz}, {Casentini}, {Caudill},
  {Cavagli{\`a}}, {Cavalier}, {Cavalieri}, {Cella}, {Cepeda}, {Cerboni
  Baiardi}, {Cerretani}, {Cesarini}, {Chamberlin}, {Chan}, {Chao}, {Charlton},
  {Chassande-Mottin}, {Cheeseboro}, {Chen}, {Chen}, {Cheng}, {Chincarini},
  {Chiummo}, {Cho}, {Cho}, {Chow}, {Christensen}, {Chu}, {Chua}, {Chung},
  {Ciani}, {Clara}, {Clark}, {Cleva}, {Coccia}, {Cohadon}, {Colla}, {Collette},
  {Cominsky}, {Constancio}, {Conte}, {Conti}, {Cook}, {Corbitt}, {Cornish},
  {Corsi}, {Cortese}, {Costa}, {Coughlin}, {Coughlin}, {Coulon}, {Countryman},
  {Couvares}, {Cowan}, {Coward}, {Cowart}, {Coyne}, {Coyne}, {Craig},
  {Creighton}, {Cripe}, {Crowder}, {Cumming}, {Cunningham}, {Cuoco}, {Dal
  Canton}, {Danilishin}, {D'Antonio}, {Danzmann}, {Darman}, {Dasgupta}, {Da
  Silva Costa}, {Dattilo}, {Dave}, {Davier}, {Davies}, {Daw}, {Day}, {De},
  {DeBra}, {Debreczeni}, {Degallaix}, {De Laurentis}, {Del{\'e}glise}, {Del
  Pozzo}, {Denker}, {Dent}, {Dergachev}, {De Rosa}, {DeRosa}, {DeSalvo},
  {Devine}, {Dhurand har}, {D{\'\i}az}, {Di Fiore}, {Di Giovanni}, {Di
  Girolamo}, {Di Lieto}, {Di Pace}, {Di Palma}, {Di Virgilio}, {Dolique},
  {Donovan}, {Dooley}, {Doravari}, {Douglas}, {Downes}, {Drago}, {Drever},
  {Driggers}, {Ducrot}, {Dwyer}, {Edo}, {Edwards}, {Effler}, {Eggenstein},
  {Ehrens}, {Eichholz}, {Eikenberry}, {Engels}, {Essick}, {Etzel}, {Evans},
  {Evans}, {Everett}, {Factourovich}, {Fafone}, {Fair}, {Fairhurst}, {Fan},
  {Fang}, {Farinon}, {Farr}, {Farr}, {Favata}, {Fays}, {Fehrmann}, {Fejer},
  {Fenyvesi}, {Ferrante}, {Ferreira}, {Ferrini}, {Fidecaro}, {Fiori},
  {Fiorucci}, {Fisher}, {Flaminio}, {Fletcher}, {Fong}, {Fournier}, {Frasca},
  {Frasconi}, {Frei}, {Freise}, {Frey}, {Frey}, {Fritschel}, {Frolov}, {Fulda},
  {Fyffe}, {Gabbard}, {Gaebel}, {Gair}, {Gammaitoni}, {Gaonkar}, {Garufi},
  {Gaur}, {Gehrels}, {Gemme}, {Geng}, {Genin}, {Gennai}, {George}, {Gergely},
  {Germain}, {Ghosh}, {Ghosh}, {Ghosh}, {Giaime}, {Giardina}, {Giazotto},
  {Gill}, {Glaefke}, {Goetz}, {Goetz}, {Gondan}, {Gonz{\'a}lez}, {Gonzalez
  Castro}, {Gopakumar}, {Gordon}, {Gorodetsky}, {Gossan}, {Gosselin}, {Gouaty},
  {Grado}, {Graef}, {Graff}, {Granata}, {Grant}, {Gras}, {Gray}, {Greco},
  {Green}, {Groot}, {Grote}, {Grunewald}, {Guidi}, {Guo}, {Gupta}, {Gupta},
  {Gushwa}, {Gustafson}, {Gustafson}, {Hacker}, {Hall}, {Hall}, {Hamilton},
  {Hammond}, {Haney}, {Hanke}, {Hanks}, {Hanna}, {Hannam}, {Hanson},
  {Hardwick}, {Harms}, {Harry}, {Harry}, {Hart}, {Hartman}, {Haster},
  {Haughian}, {Healy}, {Heidmann}, {Heintze}, {Heitmann}, {Hello}, {Hemming},
  {Hendry}, {Heng}, {Hennig}, {Henry}, {Heptonstall}, {Heurs}, {Hild}, {Hoak},
  {Hofman}, {Holt}, {Holz}, {Hopkins}, {Hough}, {Houston}, {Howell}, {Hu},
  {Huang}, {Huerta}, {Huet}, {Hughey}, {Husa}, {Huttner}, {Huynh-Dinh},
  {Indik}, {Ingram}, {Inta}, {Isa}, {Isac}, {Isi}, {Isogai}, {Iyer}, {Izumi},
  {Jacqmin}, {Jang}, {Jani}, {Jaranowski}, {Jawahar}, {Jian},
  {Jim{\'e}nez-Forteza}, {Johnson}, {Johnson-McDaniel}, {Jones}, {Jones},
  {Jonker}, {Ju}, {K}, {Kalaghatgi}, {Kalogera}, {Kandhasamy}, {Kang},
  {Kanner}, {Kapadia}, {Karki}, {Karvinen}, {Kasprzack}, {Katsavounidis},
  {Katzman}, {Kaufer}, {Kaur}, {Kawabe}, {K{\'e}f{\'e}lian}, {Kehl}, {Keitel},
  {Kelley}, {Kells}, {Kennedy}, {Key}, {Khalili}, {Khan}, {Khan}, {Khan},
  {Khazanov}, {Kijbunchoo}, {Kim}, {Kim}, {Kim}, {Kim}, {Kim}, {Kim}, {Kim},
  {Kimbrell}, {King}, {King}, {Kissel}, {Klein}, {Kleybolte}, {Klimenko},
  {Koehlenbeck}, {Koley}, {Kondrashov}, {Kontos}, {Korobko}, {Korth},
  {Kowalska}, {Kozak}, {Kringel}, {Krishnan}, {Kr{\'o}lak}, {Krueger}, {Kuehn},
  {Kumar}, {Kumar}, {Kuo}, {Kutynia}, {Lackey}, {Land ry}, {Lange}, {Lantz},
  {Lasky}, {Laxen}, {Lazzarini}, {Lazzaro}, {Leaci}, {Leavey}, {Lebigot},
  {Lee}, {Lee}, {Lee}, {Lee}, {Lenon}, {Leonardi}, {Leong}, {Leroy},
  {Letendre}, {Levin}, {Lewis}, {Li}, {Libson}, {Littenberg}, {Lockerbie},
  {Lombardi}, {London}, {Lord}, {Lorenzini}, {Loriette}, {Lormand}, {Losurdo},
  {Lough}, {Lousto}, {L{\"u}ck}, {Lundgren}, {Lynch}, {Ma}, {Machenschalk},
  {MacInnis}, {Macleod}, {Maga{\~n}a-Sandoval}, {Maga{\~n}a Zertuche}, {Magee},
  {Majorana}, {Maksimovic}, {Malvezzi}, {Man}, {Mandel}, {Mandic}, {Mangano},
  {Mansell}, {Manske}, {Mantovani}, {Marchesoni}, {Marion}, {M{\'a}rka},
  {M{\'a}rka}, {Markosyan}, {Maros}, {Martelli}, {Martellini}, {Martin},
  {Martynov}, {Marx}, {Mason}, {Masserot}, {Massinger}, {Masso-Reid},
  {Mastrogiovanni}, {Matichard}, {Matone}, {Mavalvala}, {Mazumder}, {McCarthy},
  {McClelland}, {McCormick}, {McGuire}, {McIntyre}, {McIver}, {McManus},
  {McRae}, {McWilliams}, {Meacher}, {Meadors}, {Meidam}, {Melatos}, {Mendell},
  {Mercer}, {Merilh}, {Merzougui}, {Meshkov}, {Messenger}, {Messick},
  {Metzdorff}, {Meyers}, {Mezzani}, {Miao}, {Michel}, {Middleton}, {Mikhailov},
  {Milano}, {Miller}, {Miller}, {Miller}, {Miller}, {Millhouse}, {Minenkov},
  {Ming}, {Mirshekari}, {Mishra}, {Mitra}, {Mitrofanov}, {Mitselmakher},
  {Mittleman}, {Moggi}, {Mohan}, {Mohapatra}, {Montani}, {Moore}, {Moore},
  {Moraru}, {Moreno}, {Morriss}, {Mossavi}, {Mours}, {Mow-Lowry}, {Mueller},
  {Muir}, {Mukherjee}, {Mukherjee}, {Mukherjee}, {Mukund}, {Mullavey}, {Munch},
  {Murphy}, {Murray}, {Mytidis}, {Nardecchia}, {Naticchioni}, {Nayak},
  {Nedkova}, {Nelemans}, {Nelson}, {Neri}, {Neunzert}, {Newton}, {Nguyen},
  {Nielsen}, {Nissanke}, {Nitz}, {Nocera}, {Nolting}, {Normandin}, {Nuttall},
  {Oberling}, {Ochsner}, {O'Dell}, {Oelker}, {Ogin}, {Oh}, {Oh}, {Ohme},
  {Oliver}, {Oppermann}, {Oram}, {O'Reilly}, {O'Shaughnessy}, {Ottaway},
  {Overmier}, {Owen}, {Pai}, {Pai}, {Palamos}, {Palashov}, {Palomba},
  {Pal-Singh}, {Pan}, {Pan}, {Pankow}, {Pannarale}, {Pant}, {Paoletti},
  {Paoli}, {Papa}, {Paris}, {Parker}, {Pascucci}, {Pasqualetti}, {Passaquieti},
  {Passuello}, {Patricelli}, {Patrick}, {Pearlstone}, {Pedraza}, {Pedurand},
  {Pekowsky}, {Pele}, {Penn}, {Perreca}, {Perri}, {Pfeiffer}, {Phelps},
  {Piccinni}, {Pichot}, {Piergiovanni}, {Pierro}, {Pillant}, {Pinard}, {Pinto},
  {Pitkin}, {Poe}, {Poggiani}, {Popolizio}, {Porter}, {Post}, {Powell},
  {Prasad}, {Predoi}, {Prestegard}, {Price}, {Prijatelj}, {Principe},
  {Privitera}, {Prix}, {Prodi}, {Prokhorov}, {Puncken}, {Punturo}, {Puppo},
  {P{\"u}rrer}, {Qi}, {Qin}, {Qiu}, {Quetschke}, {Quintero}, {Quitzow-James},
  {Raab}, {Rabeling}, {Radkins}, {Raffai}, {Raja}, {Rajan}, {Rakhmanov},
  {Rapagnani}, {Raymond}, {Razzano}, {Re}, {Read}, {Reed}, {Regimbau}, {Rei},
  {Reid}, {Reitze}, {Rew}, {Reyes}, {Ricci}, {Riles}, {Rizzo}, {Robertson},
  {Robie}, {Robinet}, {Rocchi}, {Rolland}, {Rollins}, {Roma}, {Romano},
  {Romano}, {Romanov}, {Romie}, {Rosi{\'n}ska}, {Rowan}, {R{\"u}diger},
  {Ruggi}, {Ryan}, {Sachdev}, {Sadecki}, {Sadeghian}, {Sakellariadou},
  {Salconi}, {Saleem}, {Salemi}, {Samajdar}, {Sammut}, {Sanchez}, {Sandberg},
  {Sandeen}, {Sand ers}, {Sassolas}, {Sathyaprakash}, {Saulson}, {Sauter},
  {Savage}, {Sawadsky}, {Schale}, {Schilling}, {Schmidt}, {Schmidt},
  {Schnabel}, {Schofield}, {Sch{\"o}nbeck}, {Schreiber}, {Schuette}, {Schutz},
  {Scott}, {Scott}, {Sellers}, {Sengupta}, {Sentenac}, {Sequino}, {Sergeev},
  {Setyawati}, {Shaddock}, {Shaffer}, {Shahriar}, {Shaltev}, {Shapiro},
  {Shawhan}, {Sheperd}, {Shoemaker}, {Shoemaker}, {Siellez}, {Siemens},
  {Sieniawska}, {Sigg}, {Silva}, {Singer}, {Singer}, {Singh}, {Singh},
  {Singhal}, {Sintes}, {Slagmolen}, {Smith}, {Smith}, {Smith}, {Son}, {Sorazu},
  {Sorrentino}, {Souradeep}, {Srivastava}, {Staley}, {Steinke}, {Steinlechner},
  {Steinlechner}, {Steinmeyer}, {Stephens}, {Stevenson}, {Stone}, {Strain},
  {Straniero}, {Stratta}, {Strauss}, {Strigin}, {Sturani}, {Stuver},
  {Summerscales}, {Sun}, {Sunil}, {Sutton}, {Swinkels}, {Szczepa{\'n}czyk},
  {Tacca}, {Talukder}, {Tanner}, {T{\'a}pai}, {Tarabrin}, {Taracchini},
  {Taylor}, {Theeg}, {Thirugnanasamband am}, {Thomas}, {Thomas}, {Thomas},
  {Thorne}, {Thrane}, {Tiwari}, {Tiwari}, {Tokmakov}, {Toland}, {Tomlinson},
  {Tonelli}, {Tornasi}, {Torres}, {Torrie}, {T{\"o}yr{\"a}}, {Travasso},
  {Traylor}, {Trifir{\`o}}, {Tringali}, {Trozzo}, {Tse}, {Turconi},
  {Tuyenbayev}, {Ugolini}, {Unnikrishnan}, {Urban}, {Usman}, {Vahlbruch},
  {Vajente}, {Valdes}, {Vallisneri}, {van Bakel}, {van Beuzekom}, {van den
  Brand}, {Van Den Broeck}, {Vand er-Hyde}, {van der Schaaf}, {van Heijningen},
  {van Veggel}, {Vardaro}, {Vass}, {Vas{\'u}th}, {Vaulin}, {Vecchio},
  {Vedovato}, {Veitch}, {Veitch}, {Venkateswara}, {Verkindt}, {Vetrano},
  {Vicer{\'e}}, {Vinciguerra}, {Vine}, {Vinet}, {Vitale}, {Vo}, {Vocca},
  {Vorvick}, {Voss}, {Vousden}, {Vyatchanin}, {Wade}, {Wade}, {Wade}, {Walker},
  {Wallace}, {Walsh}, {Wang}, {Wang}, {Wang}, {Wang}, {Wang}, {Ward}, {Warner},
  {Was}, {Weaver}, {Wei}, {Weinert}, {Weinstein}, {Weiss}, {Wen}, {We{\ss}els},
  {Westphal}, {Wette}, {Whelan}, {Whitcomb}, {Whiting}, {Williams},
  {Williamson}, {Willis}, {Willke}, {Wimmer}, {Winkler}, {Wipf}, {Wittel},
  {Woan}, {Woehler}, {Worden}, {Wright}, {Wu}, {Wu}, {Yablon}, {Yam},
  {Yamamoto}, {Yancey}, {Yu}, {Yvert}, {Zadro{\.Z}ny}, {Zangrando}, {Zanolin},
  {Zendri}, {Zevin}, {Zhang}, {Zhang}, {Zhang}, {Zhao}, {Zhou}, {Zhou}, {Zhu},
  {Zucker}, {Zuraw}, {Zweizig}, {LIGO Scientific Collaboration}, \& {Virgo
  Collaboration}}]{2016PhRvX...6d1015A}
---. 2016{\natexlab{a}}, Phys. Rev. X, 6, 041015

\bibitem[{{Abbott} {et~al.}(2016{\natexlab{b}}){Abbott}, {Abbott}, {Abbott},
  {Abernathy}, {Acernese}, {Ackley}, {Adams}, {Adams}, {Addesso}, {Adhikari},
  {Adya}, {Affeldt}, {Agathos}, {Agatsuma}, {Aggarwal}, {Aguiar}, {Aiello},
  {Ain}, {Ajith}, {Allen}, {Allocca}, {Altin}, {Anderson}, {Anderson}, {Arai},
  {Araya}, {Arceneaux}, {Areeda}, {Arnaud}, {Arun}, {Ascenzi}, {Ashton}, {Ast},
  {Aston}, {Astone}, {Aufmuth}, {Aulbert}, {Babak}, {Bacon}, {Bader}, {Baker},
  {Baldaccini}, {Ballardin}, {Ballmer}, {Barayoga}, {Barclay}, {Barish},
  {Barker}, {Barone}, {Barr}, {Barsotti}, {Barsuglia}, {Barta}, {Bartlett},
  {Bartos}, {Bassiri}, {Basti}, {Batch}, {Baune}, {Bavigadda}, {Bazzan},
  {Behnke}, {Bejger}, {Bell}, {Bell}, {Berger}, {Bergman}, {Bergmann}, {Berry},
  {Bersanetti}, {Bertolini}, {Betzwieser}, {Bhagwat}, {Bhand are}, {Bilenko},
  {Billingsley}, {Birch}, {Birney}, {Biscans}, {Bisht}, {Bitossi}, {Biwer},
  {Bizouard}, {Blackburn}, {Blair}, {Blair}, {Blair}, {Bloemen}, {Bock},
  {Bodiya}, {Boer}, {Bogaert}, {Bogan}, {Bohe}, {Bojtos}, {Bond}, {Bondu},
  {Bonnand}, {Boom}, {Bork}, {Boschi}, {Bose}, {Bouffanais}, {Bozzi},
  {Bradaschia}, {Brady}, {Braginsky}, {Branchesi}, {Brau}, {Briant}, {Brillet},
  {Brinkmann}, {Brisson}, {Brockill}, {Brooks}, {Brown}, {Brown}, {Brown},
  {Buchanan}, {Buikema}, {Bulik}, {Bulten}, {Buonanno}, {Buskulic}, {Buy},
  {Byer}, {Cadonati}, {Cagnoli}, {Cahillane}, {Calder{\'o}n Bustillo},
  {Callister}, {Calloni}, {Camp}, {Cannon}, {Cao}, {Capano}, {Capocasa},
  {Carbognani}, {Caride}, {Casanueva Diaz}, {Casentini}, {Caudill},
  {Cavagli{\`a}}, {Cavalier}, {Cavalieri}, {Cella}, {Cepeda}, {Cerboni
  Baiardi}, {Cerretani}, {Cesarini}, {Chakraborty}, {Chalermsongsak},
  {Chamberlin}, {Chan}, {Chao}, {Charlton}, {Chassande-Mottin}, {Chen}, {Chen},
  {Cheng}, {Chincarini}, {Chiummo}, {Cho}, {Cho}, {Chow}, {Christensen}, {Chu},
  {Chua}, {Chung}, {Ciani}, {Clara}, {Clark}, {Cleva}, {Coccia}, {Cohadon},
  {Colla}, {Collette}, {Cominsky}, {Constancio}, {Conte}, {Conti}, {Cook},
  {Corbitt}, {Cornish}, {Corsi}, {Cortese}, {Costa}, {Coughlin}, {Coughlin},
  {Coulon}, {Countryman}, {Couvares}, {Cowan}, {Coward}, {Cowart}, {Coyne},
  {Coyne}, {Craig}, {Creighton}, {Cripe}, {Crowder}, {Cumming}, {Cunningham},
  {Cuoco}, {Dal Canton}, {Danilishin}, {D'Antonio}, {Danzmann}, {Darman},
  {Dattilo}, {Dave}, {Daveloza}, {Davier}, {Davies}, {Daw}, {Day}, {De},
  {DeBra}, {Debreczeni}, {Degallaix}, {De Laurentis}, {Del{\'e}glise}, {Del
  Pozzo}, {Denker}, {Dent}, {Dereli}, {Dergachev}, {De Rosa}, {DeRosa},
  {DeSalvo}, {Dhurand har}, {D{\'\i}az}, {Di Fiore}, {Di Giovanni}, {Di Lieto},
  {Di Pace}, {Di Palma}, {Di Virgilio}, {Dojcinoski}, {Dolique}, {Donovan},
  {Dooley}, {Doravari}, {Douglas}, {Downes}, {Drago}, {Drever}, {Driggers},
  {Du}, {Ducrot}, {Dwyer}, {Edo}, {Edwards}, {Effler}, {Eggenstein}, {Ehrens},
  {Eichholz}, {Eikenberry}, {Engels}, {Essick}, {Etzel}, {Evans}, {Evans},
  {Everett}, {Factourovich}, {Fafone}, {Fair}, {Fairhurst}, {Fan}, {Fang},
  {Farinon}, {Farr}, {Farr}, {Favata}, {Fays}, {Fehrmann}, {Fejer}, {Ferrante},
  {Ferreira}, {Ferrini}, {Fidecaro}, {Fiori}, {Fiorucci}, {Fisher}, {Flaminio},
  {Fletcher}, {Fong}, {Fournier}, {Franco}, {Frasca}, {Frasconi}, {Frei},
  {Freise}, {Frey}, {Frey}, {Fricke}, {Fritschel}, {Frolov}, {Fulda}, {Fyffe},
  {Gabbard}, {Gair}, {Gammaitoni}, {Gaonkar}, {Garufi}, {Gatto}, {Gaur},
  {Gehrels}, {Gemme}, {Gendre}, {Genin}, {Gennai}, {George}, {Gergely},
  {Germain}, {Ghosh}, {Ghosh}, {Giaime}, {Giardina}, {Giazotto}, {Gill},
  {Glaefke}, {Goetz}, {Goetz}, {Gondan}, {Gonz{\'a}lez}, {Gonzalez Castro},
  {Gopakumar}, {Gordon}, {Gorodetsky}, {Gossan}, {Gosselin}, {Gouaty}, {Graef},
  {Graff}, {Granata}, {Grant}, {Gras}, {Gray}, {Greco}, {Green}, {Groot},
  {Grote}, {Grunewald}, {Guidi}, {Guo}, {Gupta}, {Gupta}, {Gushwa},
  {Gustafson}, {Gustafson}, {Hacker}, {Hall}, {Hall}, {Hammond}, {Haney},
  {Hanke}, {Hanks}, {Hanna}, {Hannam}, {Hanson}, {Hardwick}, {Harms}, {Harry},
  {Harry}, {Hart}, {Hartman}, {Haster}, {Haughian}, {Heidmann}, {Heintze},
  {Heitmann}, {Hello}, {Hemming}, {Hendry}, {Heng}, {Hennig}, {Heptonstall},
  {Heurs}, {Hild}, {Hoak}, {Hodge}, {Hofman}, {Hollitt}, {Holt}, {Holz},
  {Hopkins}, {Hosken}, {Hough}, {Houston}, {Howell}, {Hu}, {Huang}, {Huerta},
  {Huet}, {Hughey}, {Husa}, {Huttner}, {Huynh-Dinh}, {Idrisy}, {Indik},
  {Ingram}, {Inta}, {Isa}, {Isac}, {Isi}, {Islas}, {Isogai}, {Iyer}, {Izumi},
  {Jacqmin}, {Jang}, {Jani}, {Jaranowski}, {Jawahar}, {Jim{\'e}nez-Forteza},
  {Johnson}, {Jones}, {Jones}, {Jonker}, {Ju}, {K}, {Kalaghatgi}, {Kalogera},
  {Kandhasamy}, {Kang}, {Kanner}, {Karki}, {Kasprzack}, {Katsavounidis},
  {Katzman}, {Kaufer}, {Kaur}, {Kawabe}, {Kawazoe}, {K{\'e}f{\'e}lian}, {Kehl},
  {Keitel}, {Kelley}, {Kells}, {Kennedy}, {Key}, {Khalaidovski}, {Khalili},
  {Khan}, {Khan}, {Khan}, {Khazanov}, {Kijbunchoo}, {Kim}, {Kim}, {Kim}, {Kim},
  {Kim}, {Kim}, {King}, {King}, {Kinzel}, {Kissel}, {Kleybolte}, {Klimenko},
  {Koehlenbeck}, {Kokeyama}, {Koley}, {Kondrashov}, {Kontos}, {Korobko},
  {Korth}, {Kowalska}, {Kozak}, {Kringel}, {Krishnan}, {Kr{\'o}lak}, {Krueger},
  {Kuehn}, {Kumar}, {Kuo}, {Kutynia}, {Lackey}, {Landry}, {Lange}, {Lantz},
  {Lasky}, {Lazzarini}, {Lazzaro}, {Leaci}, {Leavey}, {Lebigot}, {Lee}, {Lee},
  {Lee}, {Lee}, {Lenon}, {Leonardi}, {Leong}, {Leroy}, {Letendre}, {Levin},
  {Levine}, {Li}, {Libson}, {Littenberg}, {Lockerbie}, {Logue}, {Lombardi},
  {Lord}, {Lorenzini}, {Loriette}, {Lormand}, {Losurdo}, {Lough}, {L{\"u}ck},
  {Lundgren}, {Luo}, {Lynch}, {Ma}, {MacDonald}, {Machenschalk}, {MacInnis},
  {Macleod}, {Maga{\~n}a-Sandoval}, {Magee}, {Mageswaran}, {Majorana},
  {Maksimovic}, {Malvezzi}, {Man}, {Mandel}, {Mandic}, {Mangano}, {Mansell},
  {Manske}, {Mantovani}, {Marchesoni}, {Marion}, {M{\'a}rka}, {M{\'a}rka},
  {Markosyan}, {Maros}, {Martelli}, {Martellini}, {Martin}, {Martin},
  {Martynov}, {Marx}, {Mason}, {Masserot}, {Massinger}, {Masso-Reid},
  {Matichard}, {Matone}, {Mavalvala}, {Mazumder}, {Mazzolo}, {McCarthy},
  {McClelland}, {McCormick}, {McGuire}, {McIntyre}, {McIver}, {McManus},
  {McWilliams}, {Meacher}, {Meadors}, {Meidam}, {Melatos}, {Mendell},
  {Mendoza-Gandara}, {Mercer}, {Merilh}, {Merzougui}, {Meshkov}, {Messenger},
  {Messick}, {Meyers}, {Mezzani}, {Miao}, {Michel}, {Middleton}, {Mikhailov},
  {Milano}, {Miller}, {Millhouse}, {Minenkov}, {Ming}, {Mirshekari}, {Mishra},
  {Mitra}, {Mitrofanov}, {Mitselmakher}, {Mittleman}, {Moggi}, {Mohan},
  {Mohapatra}, {Montani}, {Moore}, {Moore}, {Moraru}, {Moreno}, {Morriss},
  {Mossavi}, {Mours}, {Mow-Lowry}, {Mueller}, {Mueller}, {Muir}, {Mukherjee},
  {Mukherjee}, {Mukherjee}, {Mukund}, {Mullavey}, {Munch}, {Murphy}, {Murray},
  {Mytidis}, {Nardecchia}, {Naticchioni}, {Nayak}, {Necula}, {Nedkova},
  {Nelemans}, {Neri}, {Neunzert}, {Newton}, {Nguyen}, {Nielsen}, {Nissanke},
  {Nitz}, {Nocera}, {Nolting}, {Normandin}, {Nuttall}, {Oberling}, {Ochsner},
  {O'Dell}, {Oelker}, {Ogin}, {Oh}, {Oh}, {Ohme}, {Oliver}, {Oppermann},
  {Oram}, {O'Reilly}, {O'Shaughnessy}, {Ottaway}, {Ottens}, {Overmier}, {Owen},
  {Pai}, {Pai}, {Palamos}, {Palashov}, {Palomba}, {Pal-Singh}, {Pan}, {Pankow},
  {Pannarale}, {Pant}, {Paoletti}, {Paoli}, {Papa}, {Paris}, {Parker},
  {Pascucci}, {Pasqualetti}, {Passaquieti}, {Passuello}, {Patricelli},
  {Patrick}, {Pearlstone}, {Pedraza}, {Pedurand}, {Pekowsky}, {Pele}, {Penn},
  {Perreca}, {Phelps}, {Piccinni}, {Pichot}, {Piergiovanni}, {Pierro},
  {Pillant}, {Pinard}, {Pinto}, {Pitkin}, {Poggiani}, {Popolizio}, {Porter},
  {Post}, {Powell}, {Prasad}, {Predoi}, {Premachandra}, {Prestegard}, {Price},
  {Prijatelj}, {Principe}, {Privitera}, {Prodi}, {Prokhorov}, {Puncken},
  {Punturo}, {Puppo}, {P{\"u}rrer}, {Qi}, {Qin}, {Quetschke}, {Quintero},
  {Quitzow-James}, {Raab}, {Rabeling}, {Radkins}, {Raffai}, {Raja},
  {Rakhmanov}, {Rapagnani}, {Raymond}, {Razzano}, {Re}, {Read}, {Reed},
  {Regimbau}, {Rei}, {Reid}, {Reitze}, {Rew}, {Reyes}, {Ricci}, {Riles},
  {Robertson}, {Robie}, {Robinet}, {Rocchi}, {Rolland}, {Rollins}, {Roma},
  {Romano}, {Romanov}, {Romie}, {Rosi{\'n}ska}, {Rowan}, {R{\"u}diger},
  {Ruggi}, {Ryan}, {Sachdev}, {Sadecki}, {Sadeghian}, {Salconi}, {Saleem},
  {Salemi}, {Samajdar}, {Sammut}, {Sampson}, {Sanchez}, {Sandberg}, {Sand een},
  {Sanders}, {Sassolas}, {Sathyaprakash}, {Saulson}, {Sauter}, {Savage},
  {Sawadsky}, {Schale}, {Schilling}, {Schmidt}, {Schmidt}, {Schnabel},
  {Schofield}, {Sch{\"o}nbeck}, {Schreiber}, {Schuette}, {Schutz}, {Scott},
  {Scott}, {Sellers}, {Sengupta}, {Sentenac}, {Sequino}, {Sergeev}, {Serna},
  {Setyawati}, {Sevigny}, {Shaddock}, {Shah}, {Shahriar}, {Shaltev}, {Shao},
  {Shapiro}, {Shawhan}, {Sheperd}, {Shoemaker}, {Shoemaker}, {Siellez},
  {Siemens}, {Sigg}, {Silva}, {Simakov}, {Singer}, {Singer}, {Singh}, {Singh},
  {Singhal}, {Sintes}, {Slagmolen}, {Smith}, {Smith}, {Smith}, {Son}, {Sorazu},
  {Sorrentino}, {Souradeep}, {Srivastava}, {Staley}, {Steinke}, {Steinlechner},
  {Steinlechner}, {Steinmeyer}, {Stephens}, {Stevenson}, {Stone}, {Strain},
  {Straniero}, {Stratta}, {Strauss}, {Strigin}, {Sturani}, {Stuver},
  {Summerscales}, {Sun}, {Sutton}, {Swinkels}, {Szczepa{\'n}czyk}, {Tacca},
  {Talukder}, {Tanner}, {T{\'a}pai}, {Tarabrin}, {Taracchini}, {Taylor},
  {Theeg}, {Thirugnanasambandam}, {Thomas}, {Thomas}, {Thomas}, {Thorne},
  {Thorne}, {Thrane}, {Tiwari}, {Tiwari}, {Tokmakov}, {Tomlinson}, {Tonelli},
  {Torres}, {Torrie}, {T{\"o}yr{\"a}}, {Travasso}, {Traylor}, {Trifir{\`o}},
  {Tringali}, {Trozzo}, {Tse}, {Turconi}, {Tuyenbayev}, {Ugolini},
  {Unnikrishnan}, {Urban}, {Usman}, {Vahlbruch}, {Vajente}, {Valdes},
  {Vallisneri}, {van Bakel}, {van Beuzekom}, {van den Brand}, {Van Den Broeck},
  {Vand er-Hyde}, {van der Schaaf}, {van Heijningen}, {van Veggel}, {Vardaro},
  {Vass}, {Vas{\'u}th}, {Vaulin}, {Vecchio}, {Vedovato}, {Veitch}, {Veitch},
  {Venkateswara}, {Verkindt}, {Vetrano}, {Vicer{\'e}}, {Vinciguerra}, {Vine},
  {Vinet}, {Vitale}, {Vo}, {Vocca}, {Vorvick}, {Voss}, {Vousden}, {Vyatchanin},
  {Wade}, {Wade}, {Wade}, {Walker}, {Wallace}, {Walsh}, {Wang}, {Wang}, {Wang},
  {Wang}, {Wang}, {Ward}, {Warner}, {Was}, {Weaver}, {Wei}, {Weinert},
  {Weinstein}, {Weiss}, {Welborn}, {Wen}, {We{\ss}els}, {Westphal}, {Wette},
  {Whelan}, {White}, {Whiting}, {Williams}, {Williamson}, {Willis}, {Willke},
  {Wimmer}, {Winkler}, {Wipf}, {Wittel}, {Woan}, {Worden}, {Wright}, {Wu},
  {Yablon}, {Yam}, {Yamamoto}, {Yancey}, {Yap}, {Yu}, {Yvert}, {Zadro{\.z}ny},
  {Zangrando}, {Zanolin}, {Zendri}, {Zevin}, {Zhang}, {Zhang}, {Zhang},
  {Zhang}, {Zhao}, {Zhou}, {Zhou}, {Zhu}, {Zucker}, {Zuraw}, {Zweizig}, {LIGO
  Scientific Collaboration}, \& {Virgo Collaboration}}]{2016ApJ...833L...1A}
---. 2016{\natexlab{b}}, \apj, 833, L1

\bibitem[{Acernese {et~al.}(2015)}]{TheVirgo:2014hva}
Acernese, F., {et~al.} 2015, Class. Quant. Grav., 32, 024001

\bibitem[{Anderson \& Darling(1952)}]{Anderson}
Anderson, T.~W., \& Darling, D.~A. 1952, The Annals of Mathematical Statistics,
  23, 193

\bibitem[{{Askar} {et~al.}(2017){Askar}, {Szkudlarek}, {Gondek-Rosi{\'n}ska},
  {Giersz}, \& {Bulik}}]{2017MNRAS.464L..36A}
{Askar}, A., {Szkudlarek}, M., {Gondek-Rosi{\'n}ska}, D., {Giersz}, M., \&
  {Bulik}, T. 2017, \mnras, 464, L36

\bibitem[{{Astropy Collaboration} {et~al.}(2013){Astropy Collaboration},
  {Robitaille}, {Tollerud}, {Greenfield}, {Droettboom}, {Bray}, {Aldcroft},
  {Davis}, {Ginsburg}, {Price-Whelan}, {Kerzendorf}, {Conley}, {Crighton},
  {Barbary}, {Muna}, {Ferguson}, {Grollier}, {Parikh}, {Nair}, {Unther},
  {Deil}, {Woillez}, {Conseil}, {Kramer}, {Turner}, {Singer}, {Fox}, {Weaver},
  {Zabalza}, {Edwards}, {Azalee Bostroem}, {Burke}, {Casey}, {Crawford},
  {Dencheva}, {Ely}, {Jenness}, {Labrie}, {Lim}, {Pierfederici}, {Pontzen},
  {Ptak}, {Refsdal}, {Servillat}, \& {Streicher}}]{astropy}
{Astropy Collaboration}, {Robitaille}, T.~P., {Tollerud}, E.~J., {et~al.} 2013,
  \aap, 558, A33

\bibitem[{{Belczynski} {et~al.}(2016{\natexlab{a}}){Belczynski}, {Holz},
  {Bulik}, \& {O'Shaughnessy}}]{2016Natur.534..512B}
{Belczynski}, K., {Holz}, D.~E., {Bulik}, T., \& {O'Shaughnessy}, R.
  2016{\natexlab{a}}, \nat, 534, 512

\bibitem[{{Belczynski} {et~al.}(2016{\natexlab{b}}){Belczynski}, {Repetto},
  {Holz}, {O'Shaughnessy}, {Bulik}, {Berti}, {Fryer}, \&
  {Dominik}}]{2016ApJ...819..108B}
{Belczynski}, K., {Repetto}, S., {Holz}, D.~E., {et~al.} 2016{\natexlab{b}},
  \apj, 819, 108

\bibitem[{{Bird} {et~al.}(2016){Bird}, {Cholis}, {Mu{\~n}oz},
  {Ali-Ha{\"\i}moud}, {Kamionkowski}, {Kovetz}, {Raccanelli}, \&
  {Riess}}]{2016PhRvL.116t1301B}
{Bird}, S., {Cholis}, I., {Mu{\~n}oz}, J.~B., {et~al.} 2016, \prl, 116, 201301

\bibitem[{{Bouffanais} {et~al.}(2019){Bouffanais}, {Mapelli}, {Gerosa}, {Di
  Carlo}, {Giacobbo}, {Berti}, \& {Baibhav}}]{2019arXiv190511054B}
{Bouffanais}, Y., {Mapelli}, M., {Gerosa}, D., {et~al.} 2019, arXiv e-prints,
  arXiv:1905.11054

\bibitem[{{Broadhurst} {et~al.}(2018){Broadhurst}, {Diego}, \&
  {Smoot}}]{2018arXiv180205273B}
{Broadhurst}, T., {Diego}, J.~M., \& {Smoot}, George, I. 2018, arXiv e-prints,
  arXiv:1802.05273

\bibitem[{{Cao} {et~al.}(2014){Cao}, {Li}, \& {Wang}}]{2014PhRvD..90f2003C}
{Cao}, Z., {Li}, L.-F., \& {Wang}, Y. 2014, \prd, 90, 062003

\bibitem[{{Chatterjee} {et~al.}(2017){Chatterjee}, {Rodriguez}, {Kalogera}, \&
  {Rasio}}]{2017ApJ...836L..26C}
{Chatterjee}, S., {Rodriguez}, C.~L., {Kalogera}, V., \& {Rasio}, F.~A. 2017,
  \apj, 836, L26

\bibitem[{{Chatziioannou} {et~al.}(2019){Chatziioannou}, {Cotesta}, {Ghonge},
  {Lange}, {Ng}, {Bustillo}, {Clark}, {Haster}, {Khan}, {Puerrer}, {Raymond},
  {Vitale}, {Afshari}, {Babak}, {Barkett}, {Blackman}, {Bohe}, {Boyle},
  {Buonanno}, {Campanelli}, {Carullo}, {Chu}, {Flynn}, {Fong}, {Garcia},
  {Giesler}, {Haney}, {Hannam}, {Harry}, {Healy}, {Hemberger}, {Hinder},
  {Jani}, {Khamersa}, {Kidder}, {Kumar}, {Laguna}, {Lousto}, {Lovelace},
  {Littenberg}, {London}, {Millhouse}, {Nuttall}, {Ohme}, {O'Shaughnessy},
  {Ossokine}, {Pannarale}, {Schmidt}, {Pfeiffer}, {Scheel}, {Shao},
  {Shoemaker}, {Szilagyi}, {Taracchini}, {Teukolsky}, \&
  {Zlochower}}]{2019arXiv190306742C}
{Chatziioannou}, K., {Cotesta}, R., {Ghonge}, S., {et~al.} 2019, arXiv
  e-prints, arXiv:1903.06742

\bibitem[{{Chen} {et~al.}(2017){Chen}, {Holz}, {Miller}, {Evans}, {Vitale}, \&
  {Creighton}}]{2017arXiv170908079C}
{Chen}, H.-Y., {Holz}, D.~E., {Miller}, J., {et~al.} 2017, arXiv e-prints,
  arXiv:1709.08079

\bibitem[{{Dai} {et~al.}(2017){Dai}, {Venumadhav}, \&
  {Sigurdson}}]{2017PhRvD..95d4011D}
{Dai}, L., {Venumadhav}, T., \& {Sigurdson}, K. 2017, \prd, 95, 044011

\bibitem[{{Di Carlo} {et~al.}(2019){Di Carlo}, {Giacobbo}, {Mapelli},
  {Pasquato}, {Spera}, {Wang}, \& {Haardt}}]{2019arXiv190100863D}
{Di Carlo}, U.~N., {Giacobbo}, N., {Mapelli}, M., {et~al.} 2019, arXiv
  e-prints, arXiv:1901.00863

\bibitem[{{Dominik} {et~al.}(2015){Dominik}, {Berti}, {O'Shaughnessy},
  {Mandel}, {Belczynski}, {Fryer}, {Holz}, {Bulik}, \&
  {Pannarale}}]{2015ApJ...806..263D}
{Dominik}, M., {Berti}, E., {O'Shaughnessy}, R., {et~al.} 2015, \apj, 806, 263

\bibitem[{{Eldridge} \& {Stanway}(2016)}]{2016MNRAS.462.3302E}
{Eldridge}, J.~J., \& {Stanway}, E.~R. 2016, \mnras, 462, 3302

\bibitem[{{Farr} {et~al.}(2019){Farr}, {Fishbach}, {Ye}, \&
  {Holz}}]{2019arXiv190809084F}
{Farr}, W.~M., {Fishbach}, M., {Ye}, J., \& {Holz}, D.~E. 2019, \apjl, 883, L42

\bibitem[{{Finn} \& {Chernoff}(1993)}]{1993PhRvD..47.2198F}
{Finn}, L.~S., \& {Chernoff}, D.~F. 1993, \prd, 47, 2198

\bibitem[{{Fishbach} \& {Holz}(2017)}]{2017ApJ...851L..25F}
{Fishbach}, M., \& {Holz}, D.~E. 2017, \apj, 851, L25

\bibitem[{{Fishbach} {et~al.}(2017){Fishbach}, {Holz}, \&
  {Farr}}]{2017ApJ...840L..24F}
{Fishbach}, M., {Holz}, D.~E., \& {Farr}, B. 2017, \apj, 840, L24

\bibitem[{{Fishbach} {et~al.}(2018){Fishbach}, {Holz}, \&
  {Farr}}]{2018ApJ...863L..41F}
{Fishbach}, M., {Holz}, D.~E., \& {Farr}, W.~M. 2018, \apj, 863, L41

\bibitem[{{Foreman-Mackey} {et~al.}(2014){Foreman-Mackey}, {Hogg}, \&
  {Morton}}]{2014ApJ...795...64F}
{Foreman-Mackey}, D., {Hogg}, D.~W., \& {Morton}, T.~D. 2014, \apj, 795, 64

\bibitem[{{Garc{\'\i}a-Bellido}(2017)}]{2017JPhCS.840a2032G}
{Garc{\'\i}a-Bellido}, J. 2017, in Journal of Physics Conference Series, Vol.
  840, Journal of Physics Conference Series, 012032

\bibitem[{{Gerosa} \& {Berti}(2017)}]{2017PhRvD..95l4046G}
{Gerosa}, D., \& {Berti}, E. 2017, \prd, 95, 124046

\bibitem[{{Hannuksela} {et~al.}(2019){Hannuksela}, {Haris}, {Ng}, {Kumar},
  {Mehta}, {Keitel}, {Li}, \& {Ajith}}]{2019ApJ...874L...2H}
{Hannuksela}, O.~A., {Haris}, K., {Ng}, K.~K.~Y., {et~al.} 2019, \apjl, 874, L2

\bibitem[{{Hogg} {et~al.}(2010){Hogg}, {Myers}, \&
  {Bovy}}]{2010ApJ...725.2166H}
{Hogg}, D.~W., {Myers}, A.~D., \& {Bovy}, J. 2010, \apj, 725, 2166

\bibitem[{{Hurley} {et~al.}(2016){Hurley}, {Sippel}, {Tout}, \&
  {Aarseth}}]{2016PASA...33...36H}
{Hurley}, J.~R., {Sippel}, A.~C., {Tout}, C.~A., \& {Aarseth}, S.~J. 2016,
  \pasa, 33, e036

\bibitem[{{Khan} {et~al.}(2016){Khan}, {Husa}, {Hannam}, {Ohme}, {P{\"u}rrer},
  {Forteza}, \& {Boh{\'e}}}]{2016PhRvD..93d4007K}
{Khan}, S., {Husa}, S., {Hannam}, M., {et~al.} 2016, \prd, 93, 044007

\bibitem[{{Kimball} {et~al.}(2019){Kimball}, {Berry}, \&
  {Kalogera}}]{2019arXiv190307813K}
{Kimball}, C., {Berry}, C. P.~L., \& {Kalogera}, V. 2019, arXiv e-prints,
  arXiv:1903.07813

\bibitem[{{Kolmogorov}(1933)}]{Kolmogorov}
{Kolmogorov}, A. 1933, Giornale dell’Istituto Italiano degli Attuari, 4, 83

\bibitem[{{Kovetz} {et~al.}(2017){Kovetz}, {Cholis}, {Breysse}, \&
  {Kamionkowski}}]{2017PhRvD..95j3010K}
{Kovetz}, E.~D., {Cholis}, I., {Breysse}, P.~C., \& {Kamionkowski}, M. 2017,
  \prd, 95, 103010

\bibitem[{{Kruckow} {et~al.}(2018){Kruckow}, {Tauris}, {Langer}, {Kramer}, \&
  {Izzard}}]{2018MNRAS.481.1908K}
{Kruckow}, M.~U., {Tauris}, T.~M., {Langer}, N., {Kramer}, M., \& {Izzard},
  R.~G. 2018, \mnras, 481, 1908

\bibitem[{{Loredo}(2004)}]{2004AIPC..735..195L}
{Loredo}, T.~J. 2004, in American Institute of Physics Conference Series, Vol.
  735, American Institute of Physics Conference Series, ed. R.~{Fischer},
  R.~{Preuss}, \& U.~V. {Toussaint}, 195--206

\bibitem[{{Mandel}(2010)}]{2010PhRvD..81h4029M}
{Mandel}, I. 2010, \prd, 81, 084029

\bibitem[{{Mandel} {et~al.}(2017){Mandel}, {Farr}, {Colonna}, {Stevenson},
  {Ti{\v{n}}o}, \& {Veitch}}]{2017MNRAS.465.3254M}
{Mandel}, I., {Farr}, W.~M., {Colonna}, A., {et~al.} 2017, \mnras, 465, 3254

\bibitem[{{Mandel} {et~al.}(2019){Mandel}, {Farr}, \&
  {Gair}}]{2019MNRAS.486.1086M}
{Mandel}, I., {Farr}, W.~M., \& {Gair}, J.~R. 2019, \mnras, 486, 1086

\bibitem[{{Mandel} {et~al.}(2011){Mandel}, {Narayan}, \&
  {Kirshner}}]{2011ApJ...731..120M}
{Mandel}, K.~S., {Narayan}, G., \& {Kirshner}, R.~P. 2011, \apj, 731, 120

\bibitem[{{Mapelli}(2016)}]{2016MNRAS.459.3432M}
{Mapelli}, M. 2016, \mnras, 459, 3432

\bibitem[{Ng {et~al.}(2018)Ng, Wong, Broadhurst, \& Li}]{PhysRevD.97.023012}
Ng, K. K.~Y., Wong, K. W.~K., Broadhurst, T., \& Li, T. G.~F. 2018, Phys. Rev.
  D, 97, 023012.
\newblock \url{https://link.aps.org/doi/10.1103/PhysRevD.97.023012}

\bibitem[{{Nitz} {et~al.}(2019){Nitz}, {Dent}, {Davies}, {Kumar}, {Capano},
  {Harry}, {Mazzon}, {Nuttall}, {Lundgren}, \& {T{\'a}pai}}]{Nitz:Catalog}
{Nitz}, A.~H., {Dent}, T., {Davies}, G.~S., {et~al.} 2019, arXiv e-prints,
  arXiv:1910.05331

\bibitem[{Oguri(2018)}]{10.1093/mnras/sty2145}
Oguri, M. 2018, Monthly Notices of the Royal Astronomical Society, 480, 3842.
\newblock \url{https://doi.org/10.1093/mnras/sty2145}

\bibitem[{{Pan} {et~al.}(2014){Pan}, {Buonanno}, {Taracchini}, {Kidder},
  {Mrou{\'e}}, {Pfeiffer}, {Scheel}, \& {Szil{\'a}gyi}}]{2014PhRvD..89h4006P}
{Pan}, Y., {Buonanno}, A., {Taracchini}, A., {et~al.} 2014, \prd, 89, 084006

\bibitem[{{Planck Collaboration} {et~al.}(2016){Planck Collaboration}, {Ade},
  {Aghanim}, {Arnaud}, {Ashdown}, {Aumont}, {Baccigalupi}, {Banday},
  {Barreiro}, {Bartlett}, {Bartolo}, {Battaner}, {Battye}, {Benabed},
  {Beno{\^\i}t}, {Benoit-L{\'e}vy}, {Bernard}, {Bersanelli}, {Bielewicz},
  {Bock}, {Bonaldi}, {Bonavera}, {Bond}, {Borrill}, {Bouchet}, {Boulanger},
  {Bucher}, {Burigana}, {Butler}, {Calabrese}, {Cardoso}, {Catalano},
  {Challinor}, {Chamballu}, {Chary}, {Chiang}, {Chluba}, {Christensen},
  {Church}, {Clements}, {Colombi}, {Colombo}, {Combet}, {Coulais}, {Crill},
  {Curto}, {Cuttaia}, {Danese}, {Davies}, {Davis}, {de Bernardis}, {de Rosa},
  {de Zotti}, {Delabrouille}, {D{\'e}sert}, {Di Valentino}, {Dickinson},
  {Diego}, {Dolag}, {Dole}, {Donzelli}, {Dor{\'e}}, {Douspis}, {Ducout},
  {Dunkley}, {Dupac}, {Efstathiou}, {Elsner}, {En{\ss}lin}, {Eriksen},
  {Farhang}, {Fergusson}, {Finelli}, {Forni}, {Frailis}, {Fraisse},
  {Franceschi}, {Frejsel}, {Galeotta}, {Galli}, {Ganga}, {Gauthier}, {Gerbino},
  {Ghosh}, {Giard}, {Giraud-H{\'e}raud}, {Giusarma}, {Gjerl{\o}w},
  {Gonz{\'a}lez-Nuevo}, {G{\'o}rski}, {Gratton}, {Gregorio}, {Gruppuso},
  {Gudmundsson}, {Hamann}, {Hansen}, {Hanson}, {Harrison}, {Helou},
  {Henrot-Versill{\'e}}, {Hern{\'a}ndez-Monteagudo}, {Herranz}, {Hildebrand t},
  {Hivon}, {Hobson}, {Holmes}, {Hornstrup}, {Hovest}, {Huang}, {Huffenberger},
  {Hurier}, {Jaffe}, {Jaffe}, {Jones}, {Juvela}, {Keih{\"a}nen}, {Keskitalo},
  {Kisner}, {Kneissl}, {Knoche}, {Knox}, {Kunz}, {Kurki-Suonio}, {Lagache},
  {L{\"a}hteenm{\"a}ki}, {Lamarre}, {Lasenby}, {Lattanzi}, {Lawrence}, {Leahy},
  {Leonardi}, {Lesgourgues}, {Levrier}, {Lewis}, {Liguori}, {Lilje},
  {Linden-V{\o}rnle}, {L{\'o}pez-Caniego}, {Lubin}, {Mac{\'\i}as-P{\'e}rez},
  {Maggio}, {Maino}, {Mandolesi}, {Mangilli}, {Marchini}, {Maris}, {Martin},
  {Martinelli}, {Mart{\'\i}nez-Gonz{\'a}lez}, {Masi}, {Matarrese}, {McGehee},
  {Meinhold}, {Melchiorri}, {Melin}, {Mendes}, {Mennella}, {Migliaccio},
  {Millea}, {Mitra}, {Miville-Desch{\^e}nes}, {Moneti}, {Montier}, {Morgante},
  {Mortlock}, {Moss}, {Munshi}, {Murphy}, {Naselsky}, {Nati}, {Natoli},
  {Netterfield}, {N{\o}rgaard-Nielsen}, {Noviello}, {Novikov}, {Novikov},
  {Oxborrow}, {Paci}, {Pagano}, {Pajot}, {Paladini}, {Paoletti}, {Partridge},
  {Pasian}, {Patanchon}, {Pearson}, {Perdereau}, {Perotto}, {Perrotta},
  {Pettorino}, {Piacentini}, {Piat}, {Pierpaoli}, {Pietrobon}, {Plaszczynski},
  {Pointecouteau}, {Polenta}, {Popa}, {Pratt}, {Pr{\'e}zeau}, {Prunet},
  {Puget}, {Rachen}, {Reach}, {Rebolo}, {Reinecke}, {Remazeilles}, {Renault},
  {Renzi}, {Ristorcelli}, {Rocha}, {Rosset}, {Rossetti}, {Roudier},
  {Rouill{\'e} d'Orfeuil}, {Rowan-Robinson}, {Rubi{\~n}o-Mart{\'\i}n},
  {Rusholme}, {Said}, {Salvatelli}, {Salvati}, {Sandri}, {Santos},
  {Savelainen}, {Savini}, {Scott}, {Seiffert}, {Serra}, {Shellard}, {Spencer},
  {Spinelli}, {Stolyarov}, {Stompor}, {Sudiwala}, {Sunyaev}, {Sutton},
  {Suur-Uski}, {Sygnet}, {Tauber}, {Terenzi}, {Toffolatti}, {Tomasi},
  {Tristram}, {Trombetti}, {Tucci}, {Tuovinen}, {T{\"u}rler}, {Umana},
  {Valenziano}, {Valiviita}, {Van Tent}, {Vielva}, {Villa}, {Wade}, {Wandelt},
  {Wehus}, {White}, {White}, {Wilkinson}, {Yvon}, {Zacchei}, \&
  {Zonca}}]{2016A&A...594A..13P}
{Planck Collaboration}, {Ade}, P.~A.~R., {Aghanim}, N., {et~al.} 2016, \aap,
  594, A13

\bibitem[{{Rodriguez} {et~al.}(2018){Rodriguez}, {Amaro-Seoane}, {Chatterjee},
  \& {Rasio}}]{2018PhRvL.120o1101R}
{Rodriguez}, C.~L., {Amaro-Seoane}, P., {Chatterjee}, S., \& {Rasio}, F.~A.
  2018, \prl, 120, 151101

\bibitem[{{Rodriguez} {et~al.}(2016{\natexlab{a}}){Rodriguez}, {Haster},
  {Chatterjee}, {Kalogera}, \& {Rasio}}]{2016ApJ...824L...8R}
{Rodriguez}, C.~L., {Haster}, C.-J., {Chatterjee}, S., {Kalogera}, V., \&
  {Rasio}, F.~A. 2016{\natexlab{a}}, \apj, 824, L8

\bibitem[{{Rodriguez} {et~al.}(2016{\natexlab{b}}){Rodriguez}, {Zevin},
  {Pankow}, {Kalogera}, \& {Rasio}}]{2016ApJ...832L...2R}
{Rodriguez}, C.~L., {Zevin}, M., {Pankow}, C., {Kalogera}, V., \& {Rasio},
  F.~A. 2016{\natexlab{b}}, \apj, 832, L2

\bibitem[{{Roulet} \& {Zaldarriaga}(2019)}]{2019MNRAS.484.4216R}
{Roulet}, J., \& {Zaldarriaga}, M. 2019, \mnras, 484, 4216

\bibitem[{{Samsing}(2018)}]{2018PhRvD..97j3014S}
{Samsing}, J. 2018, \prd, 97, 103014

\bibitem[{Smirnov(1948)}]{Smirnov}
Smirnov, N. 1948, The Annals of Mathematical Statistics, 19, 279–281

\bibitem[{{Spera} {et~al.}(2019){Spera}, {Mapelli}, {Giacobbo}, {Trani},
  {Bressan}, \& {Costa}}]{2019MNRAS.485..889S}
{Spera}, M., {Mapelli}, M., {Giacobbo}, N., {et~al.} 2019, \mnras, 485, 889

\bibitem[{Stephens(1974)}]{Stephens}
Stephens, M.~A. 1974, Journal of the American Statistical Association, 69, 730

\bibitem[{{Stevenson} {et~al.}(2017{\natexlab{a}}){Stevenson}, {Berry}, \&
  {Mandel}}]{2017MNRAS.471.2801S}
{Stevenson}, S., {Berry}, C. P.~L., \& {Mandel}, I. 2017{\natexlab{a}}, \mnras,
  471, 2801

\bibitem[{{Stevenson} {et~al.}(2015){Stevenson}, {Ohme}, \&
  {Fairhurst}}]{2015ApJ...810...58S}
{Stevenson}, S., {Ohme}, F., \& {Fairhurst}, S. 2015, \apj, 810, 58

\bibitem[{{Stevenson} {et~al.}(2017{\natexlab{b}}){Stevenson},
  {Vigna-G{\'o}mez}, {Mandel}, {Barrett}, {Neijssel}, {Perkins}, \& {de
  Mink}}]{2017NatCo...814906S}
{Stevenson}, S., {Vigna-G{\'o}mez}, A., {Mandel}, I., {et~al.}
  2017{\natexlab{b}}, Nature Communications, 8, 14906

\bibitem[{{Talbot} \& {Thrane}(2018)}]{2018ApJ...856..173T}
{Talbot}, C., \& {Thrane}, E. 2018, \apj, 856, 173

\bibitem[{{Veitch} {et~al.}(2015){Veitch}, {Raymond}, {Farr}, {Farr}, {Graff},
  {Vitale}, {Aylott}, {Blackburn}, {Christensen}, {Coughlin}, {Del Pozzo},
  {Feroz}, {Gair}, {Haster}, {Kalogera}, {Littenberg}, {Mandel},
  {O'Shaughnessy}, {Pitkin}, {Rodriguez}, {R{\"o}ver}, {Sidery}, {Smith}, {Van
  Der Sluys}, {Vecchio}, {Vousden}, \& {Wade}}]{2015PhRvD..91d2003V}
{Veitch}, J., {Raymond}, V., {Farr}, B., {et~al.} 2015, \prd, 91, 042003

\bibitem[{{Venumadhav} {et~al.}(2019{\natexlab{a}}){Venumadhav}, {Zackay},
  {Roulet}, {Dai}, \& {Zaldarriaga}}]{IAS:O1}
{Venumadhav}, T., {Zackay}, B., {Roulet}, J., {Dai}, L., \& {Zaldarriaga}, M.
  2019{\natexlab{a}}, \prd, 100, 023011

\bibitem[{{Venumadhav} {et~al.}(2019{\natexlab{b}}){Venumadhav}, {Zackay},
  {Roulet}, {Dai}, \& {Zaldarriaga}}]{IAS:O2}
---. 2019{\natexlab{b}}, arXiv e-prints, arXiv:1904.07214

\bibitem[{{Vitale} {et~al.}(2017{\natexlab{a}}){Vitale}, {Lynch}, {Raymond},
  {Sturani}, {Veitch}, \& {Graff}}]{2017PhRvD..95f4053V}
{Vitale}, S., {Lynch}, R., {Raymond}, V., {et~al.} 2017{\natexlab{a}}, \prd,
  95, 064053

\bibitem[{{Vitale} {et~al.}(2017{\natexlab{b}}){Vitale}, {Lynch}, {Sturani}, \&
  {Graff}}]{2017CQGra..34cLT01V}
{Vitale}, S., {Lynch}, R., {Sturani}, R., \& {Graff}, P. 2017{\natexlab{b}},
  Classical and Quantum Gravity, 34, 03LT01

\bibitem[{{Woosley}(2016)}]{2016ApJ...824L..10W}
{Woosley}, S.~E. 2016, \apj, 824, L10

\bibitem[{{Wysocki} {et~al.}(2019){Wysocki}, {Lange}, \&
  {O'Shaughnessy}}]{2019PhRvD.100d3012W}
{Wysocki}, D., {Lange}, J., \& {O'Shaughnessy}, R. 2019, \prd, 100, 043012

\bibitem[{{Zevin} {et~al.}(2017){Zevin}, {Pankow}, {Rodriguez}, {Sampson},
  {Chase}, {Kalogera}, \& {Rasio}}]{2017ApJ...846...82Z}
{Zevin}, M., {Pankow}, C., {Rodriguez}, C.~L., {et~al.} 2017, \apj, 846, 82

\bibitem[{{Zevin} {et~al.}(2019){Zevin}, {Samsing}, {Rodriguez}, {Haster}, \&
  {Ramirez-Ruiz}}]{2019ApJ...871...91Z}
{Zevin}, M., {Samsing}, J., {Rodriguez}, C., {Haster}, C.-J., \&
  {Ramirez-Ruiz}, E. 2019, \apj, 871, 91

\end{thebibliography}
\end{document}